\documentclass[preprint,12pt]{elsarticle}

\usepackage{multicol}
\usepackage[fleqn]{amsmath}
\usepackage{subfigure}
\usepackage{bm}
\usepackage{graphicx}
\usepackage{dcolumn}
\usepackage{booktabs}
\usepackage{geometry}
\usepackage{natbib} 
\usepackage{textcomp}
\usepackage[section]{placeins}
\usepackage{threeparttable}

\journal{Nuclear Instruments and Methods in Physics Research Section A}

\begin{document}

\begin{frontmatter}

\title{Inhibition of Current-spike Formation Based on Longitudinal Phase Space Manipulation for High-Repetition-Rate X-ray FEL}

\author[1,2]{Zihan Zhu}

\author[3]{Duan Gu}

\author[4]{Jiawei Yan}

\author[3]{Zhen Wang}

\author[1,2]{Hanxiang Yang}

\author[3]{Meng Zhang}

\author[3]{Haixiao Deng}

\author[3]{Qiang Gu\corref{cor1}}
\ead{guqiang@zjlab.org.cn}
\cortext[cor1]{Corresponding author}

\address[1]{Shanghai Institute of Applied Physics, Chinese Academy of Sciences, Shanghai 201800, China}
\address[2]{University of Chinese Academy of Sciences, Beijing 100049, China}
\address[3]{Shanghai Advanced Research Institute, Chinese Academy of Sciences, Shanghai 201210, China}
\address[4]{European XFEL GmbH, 22869 Schenefeld, Germany}

\begin{abstract}
The formation of a double-horn current profile is a challenging issue in the achievement of electron bunch with high peak current, especially for high-repetition-rate X-ray free-electron lasers (XFELs) where more nonlinear and intricate beam dynamics propagation is involved. Here, we propose to correct the nonlinear beam longitudinal phase space distortion in the photoinjector section with a dual-mode buncher. In combination with the evolutionary many-objective beam dynamics optimization, this method is shown to be effective in manipulating the longitudinal phase space at the injector exit. Furthermore, the formation of the current spike is avoided after the multi-stage charge density modulation and electron bunch with a peak current of 1.6 kA is achieved for 100-pC bunch charge. Start-to-end simulations based on the Shanghai high-repetition-rate XFEL and extreme light facility demonstrate that the proposed scheme can increase the FEL pulse energy by more than 3 times in a cascading operation of echo-enabled harmonic generation and high-gain harmonic generation. Moreover, this method can also be used for longitudinal phase space shaping of electron beams operating at a high repetition rate to meet the specific demands of different researches.

\end{abstract}


\end{frontmatter}

\section{Introduction}

As a leading-edge scientific instrument, X-ray free-electron lasers (XFELs) can provide coherent X-ray pulses with high brightness and femtosecond duration \cite{Pellegrinireview,HUANG2021100097}, enabling new scientific research in physics, chemistry, biology, and materials science \cite{ostrom2015probing,kang2015crystal,dejoie2015serial,martin2016serial,son2011multiwavelength}. Over the last decade, there is a growing demand for XFELs to achieve better performance. Therefore, the beam dynamics quality of the electron beam needs to be continuously improved to satisfy the demand from various scientific communities. Moreover, many different experiments have a specific desire for the spectral and temporal structure of the X-ray pulses, which can be obtained by accurate manipulation of bunch longitudinal phase space at the entrance of the undulator.

Typically, electron beams with ultra-low transverse emittance are generated in the photoinjector and further accelerated in the linac. The magnetic chicane, consisting of four dipole magnets, is used to compress the electron bunch to obtain a high peak current. However, current horns can often be found in the current profile after strong bunch longitudinal compression, resulting in several severe problems \cite{ding2016beam,di2014electron}, especially in high-repetition-rate XFELs \cite{zhu2017sclf,yan2019multi,decking2020mhz,vogt2018status}. For instance, the current horn can generate strong coherent synchrotron radiation (CSR) that can spoil the transverse emittance, i.e. causing a nonuniform transverse tilt along the bunch, leading to the transverse misalignment between the bunch slices and projected emittance growth \cite{braun2000emittance,heifets2002coherent,di2013cancellation}. Moreover, an undesired additional longitudinal energy modulation is exerted due to the wakefield effect and longitudinal space charge (LSC) effect, both of which can largely jeopardize the FEL performance in the undulator, especially in seeded FELs which require accurate superposition between the laser and electron bunch\cite{venturini2008models,saldin2004longitudinal,huang2005microbunching}. 

Therefore, the mitigation of the undesired nonlinear density modulation in the bunch compressors is critical for linac-driven FEL radiation performance \cite{di2014electron}. In addition to the no-spike shape in the current profile, the so-called ``flat-flat" distribution in the longitudinal phase space is preferable, especially for the seeded XFELs \cite{cornacchia2006formation}. This beam property indicates that there exists a flat-top shape in the current profile as well as the flat energy distribution along with the bunch core longitudinally, which can improve the FEL performance enormously on the stable output pulse energy and facilitate the controllable spectral bandwidth \cite{bonifacio1994spectrum}.

As the cause of the current horn formation, the intense nonlinear effect consists of both the nonlinear correlated energy chirp in the bunch longitudinal phase space and the high-order longitudinal dispersion coefficients in the magnetic chicane. The nonlinear energy spread is mainly induced from the off-crest RF field curvature in the cavities, collective effects like the longitudinal wakefield and the intense space charge effect in the low-energy region where the relativistic factor $\beta$ of the bunch is not close to 1 \cite{limborg2006optimum}. For the purpose of relieving the nonlinear effect in the beam propagation through the dispersive section, many methods have been applied. The widely-used approach is the harmonic RF cavity that can linearize the longitudinal phase space by removing the quadratic correlated energy spread \cite{emma2001x}. However, the high-order parts of the longitudinal phase space correlation are left uncompensated, usually leading to the generation of current spike under the strong electron distribution modulation. A direct approach has been attempted in Linac Coherent Light Source, a collimator was utilized in the dispersive region to directly truncate the double-horn current horns at the beam head and tail \cite{ding2016beam}. However, this method can hardly be applied to high-repetition-rate XFEL facilities because of the high beam loss radiation on the collimator \cite{ding2016beam,zhou2015measurements}.

Several attempts have been conducted to compensate the longitudinal phase space for high-repetition-rate operation. The corrugated structure was placed to provide a passive wakefield for energy modulation upstream the bunch compression chicane in order to reduce the high-order energy spread in longitudinal phase space\cite{wang2018nonlinear,tnbeam}. Moreover, the nonlinear optics of the magnetic chicane can be adjusted by inserting an optics corrector like sextupole or octupole\cite{sudar2020octupole,charles2017current}, similar optical linearization method is applied to remove the quadratic energy chirp in a compact FEL design\cite{sun2014x}. Nevertheless, achievement of this target is always troublesome, especially for XFEL facility operating at the frequency of 1 MHz, which involves more complicated beam dynamics properties evolution along the beamline transmission \cite{wang2018nonlinear,gulongitudinal}.

In this work, a dual-mode buncher cavity is utilized as a new effective approach to compensating the nonlinear longitudinal phase space that mainly results from the space charge effect and the nonlinear velocity bunching in the photoinjector section. This alternate method has been adopted to the Shanghai High repetitioN rate XFEL and Extreme light facility (SHINE) injector beam dynamics design. Combined with the global optimization, the longitudinal phase space distribution at the end of linac can be improved distinctly to achieve the high peak current with no appearance of current spike under strong bunch compression. Section \ref{s2} shows the principle of the compensation method and the evolutionary many-objective beam dynamics optimization. In section \ref{s3}, the start-to-end optimization results of SHINE are presented to demonstrate improvement in electron beam quality and FEL performance. Section \ref{s4} gives the discussion about the whole research and presents a conclusion for it.

\section{Nonlinear longitudinal phase space compensation in the injector}
\label{s2} 
\subsection{Principle of the compensation method}
\label{s21} 
\begin{figure*}[htb]
	\centering
	\includegraphics[width=1\linewidth]{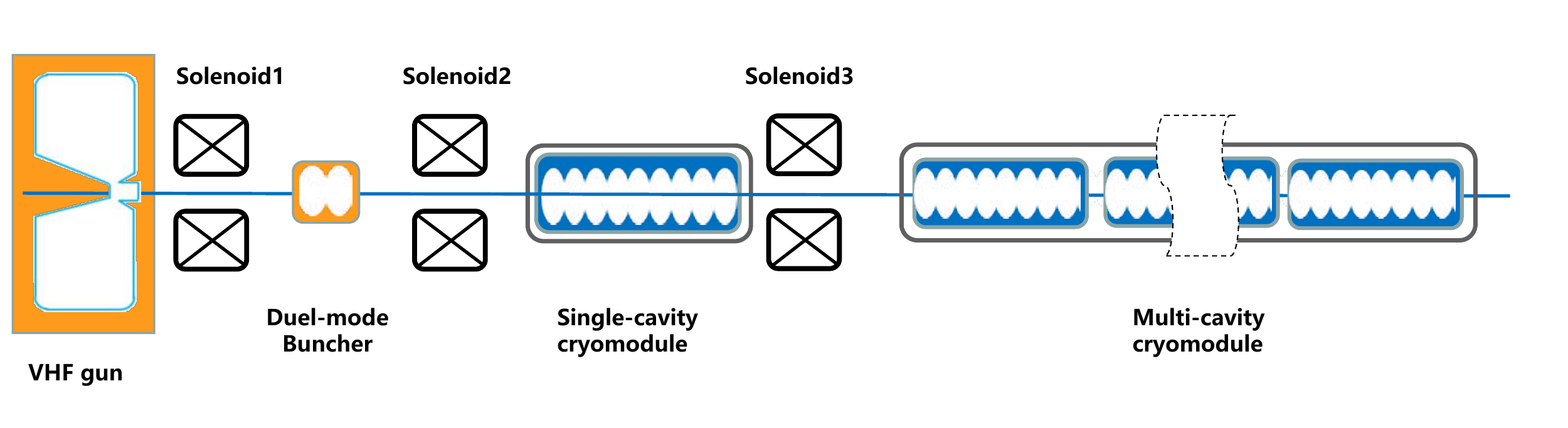}
	\caption{The conceptual schematic of the SHINE injector section. The 216.7 MHz VHF gun is followed by the dual-mode buncher, both of them are normal-conducting cavities while the single 9-cell cryomodule and eight 9-cell cryomodules are superconducting. Three sets of the solenoid are deployed along the beamline to compensate for the linear emittance growth mainly induced by RF field and space charge effect to achieve ultra-low transverse emittance.}
	\label{injector}
\end{figure*}

In this work, we first concentrate on the beam longitudinal properties improvement in the photoinjector, whose layout is shown in Fig. \ref{injector}. It is deployed to generate electron bunch with low emittance and high brightness. As the APEX-type electron gun cavity works at a relatively low frequency compared with the common rf-band gun, the beam dynamics properties evolution is intricate \cite{sannibale2019high,schmerge2014lcls,papadopoulos2012injector}. The relatively low energy of the beam at the exit of the gun cavity induces a strong and dominant space charge effect during the beam delivery, causing the nonlinear emittance growth that can not be compensated. Hence, suppression of the space charge effect is a priority for generating a bunch with low transverse emittance. Therefore, the bunch at the cathode is lengthened as the cigar regime so that some of the nonlinear space-charge effects can be eliminated \cite{bazarov2009maximum,filippetto2014maximum}. As a trade-off, the laser pulse is much longer than required for driving the FEL radiation in the undulator, which in turn necessitates a relatively larger bunch longitudinal compression ratio in the downstream beam delivery.

The velocity compression method is commonly applied in the injector which is running out of RF crest to create the energy chirp in the buncher \cite{serafini2001velocity,anderson2005velocity}. The $\beta$ = $v$/$\rm c$ cannot be treated as a perturbation in this low-energy region, which causes the differences in $\beta$ value along the bunch longitudinally. However, strong nonlinear velocity modulation combined with the nonlinear space-charge force is induced during the beam delivery, both of which lead to the nonlinearity in the beam longitudinal phase space. This results in the obvious asymmetric current profile and the nonlinear time-energy chirp, contributing to the difficulty to achieve a desirable plateau-like current profile with a high peak current at the end of the linac.  

\begin{figure}[htb] 
	\centering 
	\includegraphics[width=0.7\linewidth]{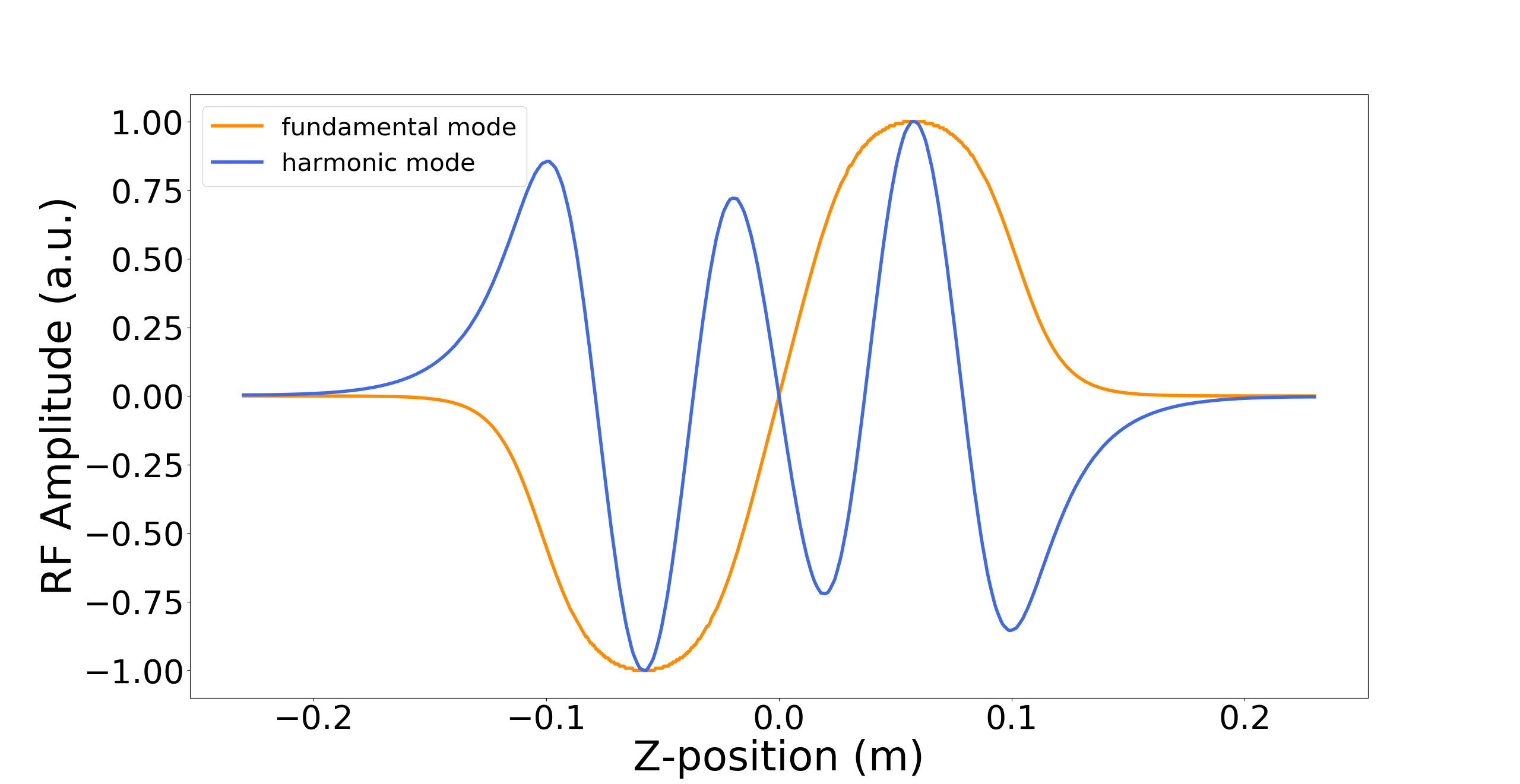}
	\caption[width=1\textwidth]{The fundamental (orange) and the harmonic (blue) electric field along the longitudinal axis in the buncher cavity.}
	\label{buncher_field} 
\end{figure}

\begin{figure}[htb] 
	\centering 
	\subfigure[]{
		\label{buncher1}	
		\includegraphics[width=0.48\linewidth]{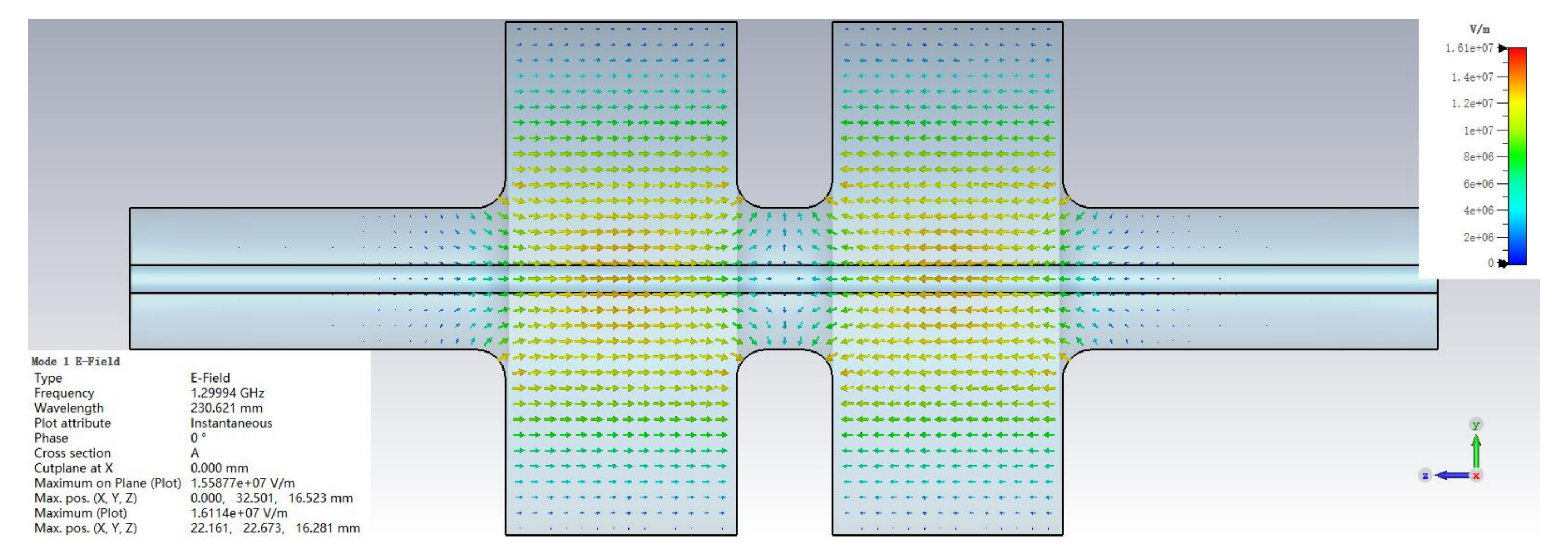}}
	\subfigure[]{
		\label{buncher2}	
		\includegraphics[width=0.48\linewidth]{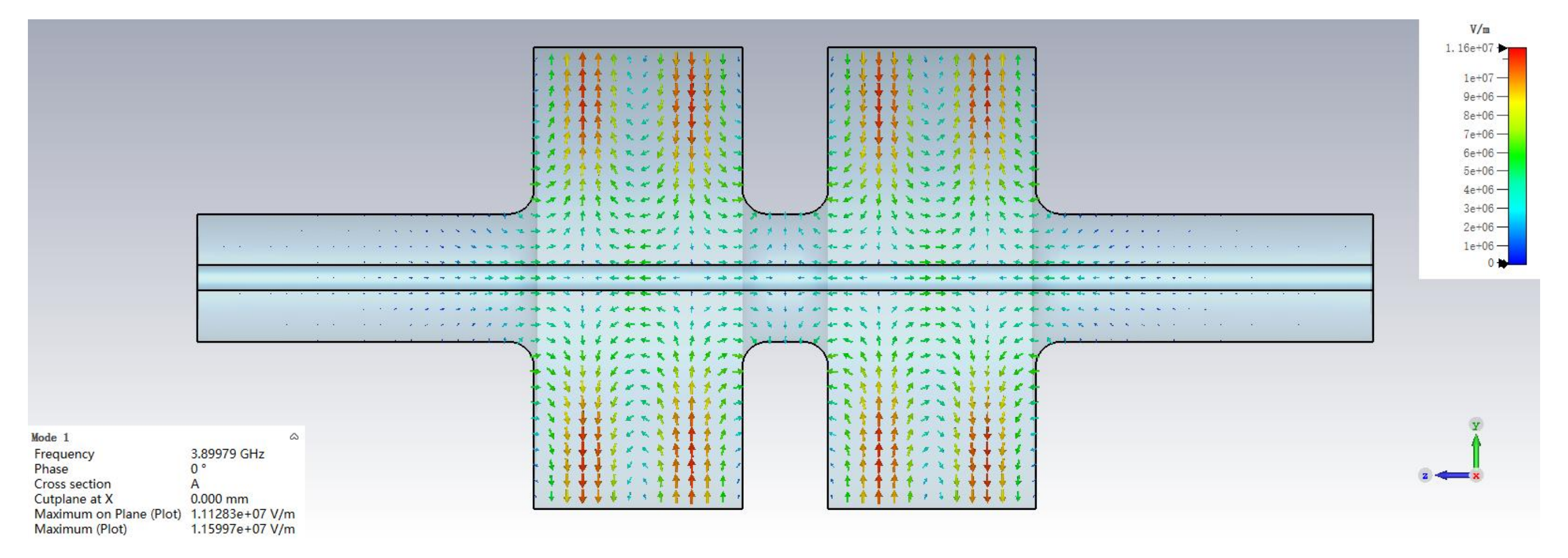}}
	\caption[width=1\textwidth]{Schematic of the electromagnetic field distribution in the cavity. (a) The 1.3 GHz fundamental mode. (b) The 3.9 GHz harmonic mode.}
	\label{buncher} 
\end{figure}

In order to solve this problem, a dual-mode buncher cavity is proposed to compensate for the RF curvature in the single fundamental mode, as is shown in Fig. \ref{buncher_field}. To be precise, the third harmonic mode is fed in the buncher to not only improve the linearity of the RF field but also compensate for the nonlinear longitudinal bunching which leads to the asymmetric current profile. The two types of RF modes will be supplied by each power source, and their voltages and phases can be set respectively. The field distribution of the dual-mode is presented in Fig. \ref{buncher}.  With the harmonic mode occupied, both the energy and density modulation in the buncher cavity are more linear, corresponding to a more symmetric current profile at the exit of the photoinjector. Furthermore, the longitudinal distribution can be adjusted by changing the amplitude and RF phase of the individual mode, which will be a more flexible and achievable approach to beam shaping at the entrance of the linac.

The photoinjector physical design involves many parameters. The amplitude and phase of each cavity are essential for the energy gain and longitudinal distribution of the bunch, and the strength and position of each solenoid set determine the precise compensation of the transverse emittance. Also, many nonlinear effects, such as velocity bunching, space-charge field, emittance compensation, and coupling between the transverse and longitudinal beam dynamics are highly correlated with each other, having a nonlinear and intricate impact on the final beam dynamics. Therefore, it can hardly get the optimal parameters with parameters scanning or adjusting manually through each simulation. Consequently, the global optimum searching algorithm is requested in the beam dynamics design.

\subsection{Many-objective photoinjector beam dynamics optimization for SHINE injector}
\label{s22}

As is mentioned above, every working point of the cavities and solenoids together with their exact positions should be designed and settled elaborately to achieve the high-quality electron bunch at the entrance of the main accelerating section. Additionally, more beam longitudinal phase space properties other than bunch length should be taken into consideration. Together with the transverse emittance, this becomes a beam dynamics optimization with more than three objectives.

The optimization with more than one objective is called a multi-objective optimization problem which has been encountered in accelerator science for a long time. In the beginning, most of these kinds of problems are solved through large-scale parameter scanning simulation to determine the final value of each control parameter. It is not only time-consuming but also difficult to find the global optimal solution. Moreover, the multi-objective problems are usually converted to single-objective ones by means of distributing each objective with corresponding weight. However, the optimal weight values are difficult to determine and this method is not feasible to study and analyze the relationships between each objective parameter. In 2005, Bazarov conducted a multi-objective optimization to the photoinjector beam dynamics design with a multi-objective evolutionary algorithm and demonstrated its efficiency and reliability \cite{bazarov2005multivariate}. From then on this method becomes a popular and typical approach in accelerator beam dynamics design \cite{bartolini2012multiobjective,gulliford2017multiobjective,qiang2019fast}. With the rapid development of computer science and high-performance cluster, many improved algorithms are accessible and the Non-dominated Sorting Genetic Algorithm III(NSGA-III) is selected to be applied to this beam dynamics optimization strategy. NSGA-III is an evolutionary many-objective optimization algorithm using the reference-point-based non-dominated sorting approach, which keeps it efficient in optimizing more than three objectives simultaneously compared with NSGA-II \cite{deb2002fast,deb2013evolutionary,yan2019generation}. Based on it, the beam dynamics simulation of the injector section is conducted using ASTRA, which is a space charge tracking algorithm widely used worldwide to simulate beam dynamics in injector sections \cite{flottmann2011astra}.

In this optimization, there exist 18 variables containing the RF phase, the amplitude of each cavity, and the detailed layout of the whole injector section. Meanwhile, the beam dynamics properties should be accurately calculated and treated as the optimization objectives. The population size of one generation is set to 400 and the total number of generations is 200. The result of the optimization algorithm is a population of solutions rather than a single one. Therefore, the accurate tradeoff and offset can be evaluated with respect to each other and lays the foundation for a more profound theoretical analysis. 

Since the final goal of this optimization is providing the electron bunch to drive the high-brightness FEL lasing, the four beam dynamics properties that are influential to the FEL performance are selected as the objectives in the optimization. Primarily, the transverse projected emittance and the bunch length are the two objectives usually settled in the previous injector optimization. However, to achieve more effective optimization and thus high-quality FEL performance, more longitudinal phase space characteristics should be taken into account. Nonlinear energy spread has a vital impact on the longitudinal compression in the magnetic chicane. Though a 3.9 GHz RF harmonic cavity is appointed as the harmonic linearizer located in the linac section that aims to mitigate the second-order correlation in the longitudinal phase space, the uncompensated third-order and above term will accumulate along with the beam delivery and result in the intense nonlinear distortion in the multi-stage chicanes. The correlated energy distribution can be fitted under the least square method as follows:
\begin{equation}
	E_{fit}=E_0+f_1z_i+f_2{z_i}^2,
\end{equation}
where $E_0$, $f_1$, and $f_2$ are the constant, first and second terms of the fitted energy distribution polynomial in the longitudinal phase space. Thus, the definition of the high-order energy spread of each macroparticle is:
\begin{equation}
	E_{H.O.}=E_z-E_{fit}=E_z-E_0-f_1z_i-f_2{z_i}^2,
\end{equation}
where $E_z$ is the actual value of energy for each macroparticle. Based on it, the root mean square (RMS) value of the high-order energy spread is defined as $\sqrt{<{E_{H.O.}^2}>}$, which is the third objective in the optimization strategy.

\begin{figure*}[htb] 
	\centering 
	\subfigure[]{
		\label{skew1}	
		\includegraphics[width=0.3\linewidth]{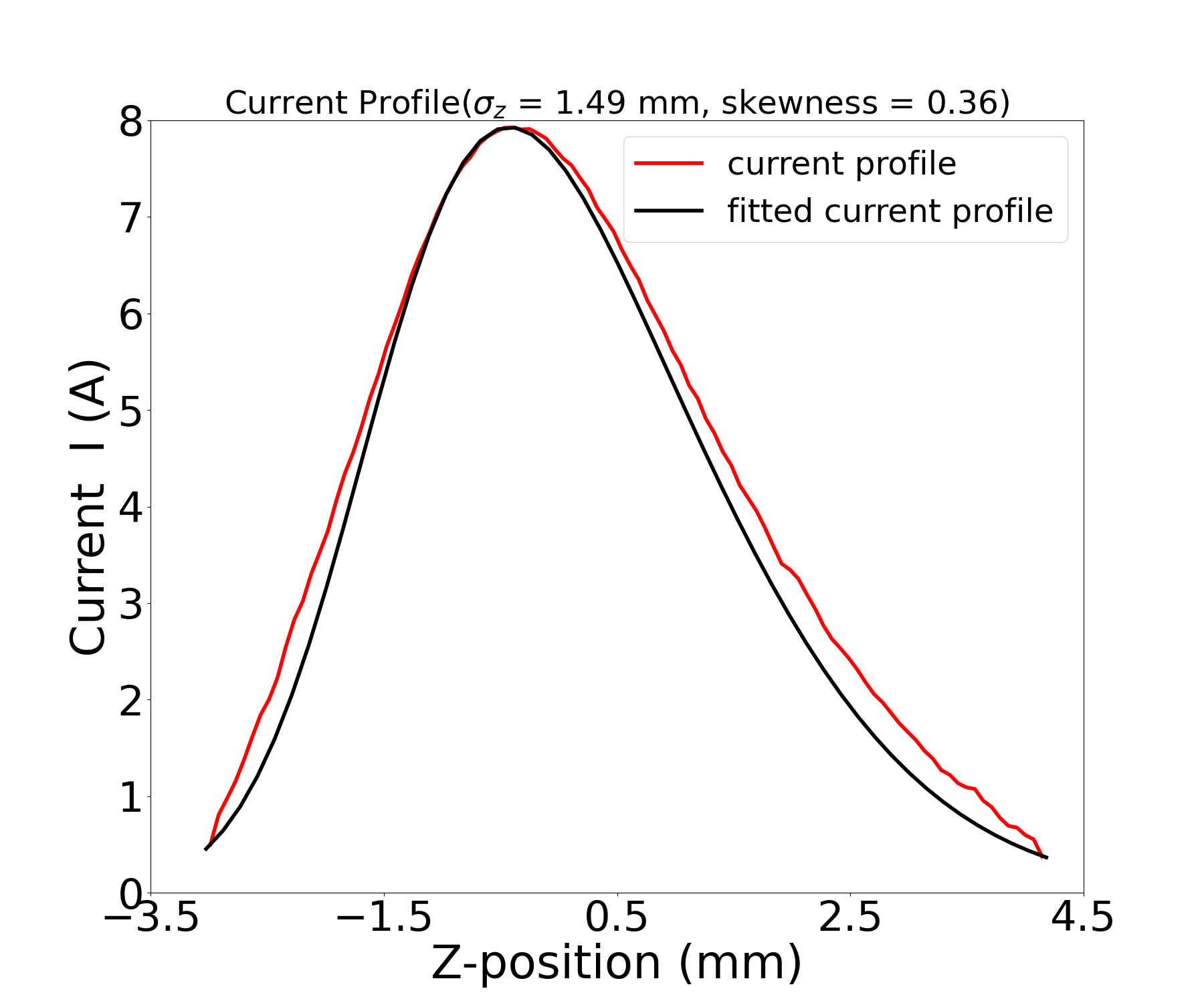}}
	\subfigure[]{
		\label{skew2}	
		\includegraphics[width=0.3\linewidth]{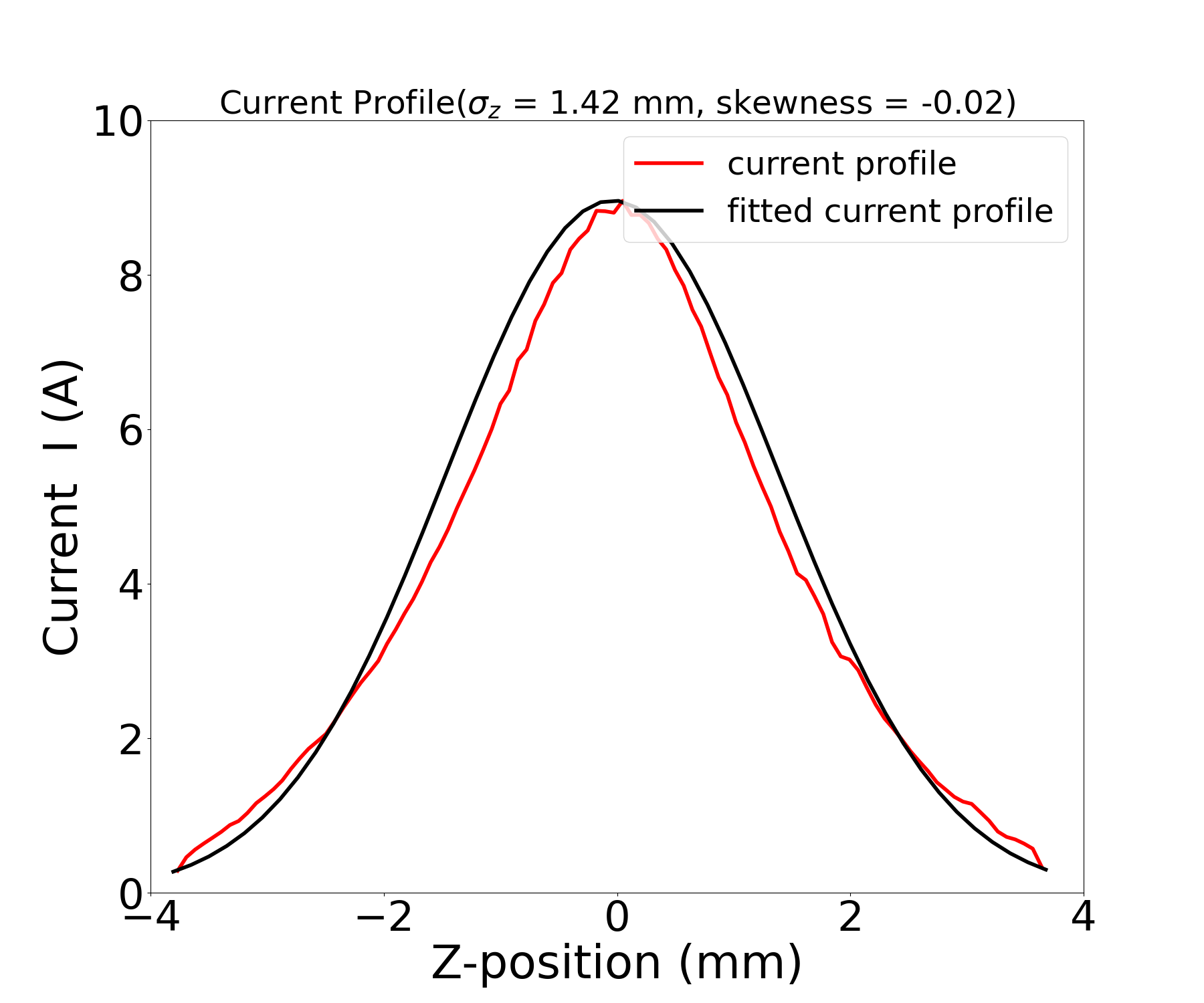}}
	\subfigure[]{
	\label{skew3}	
	\includegraphics[width=0.3\linewidth]{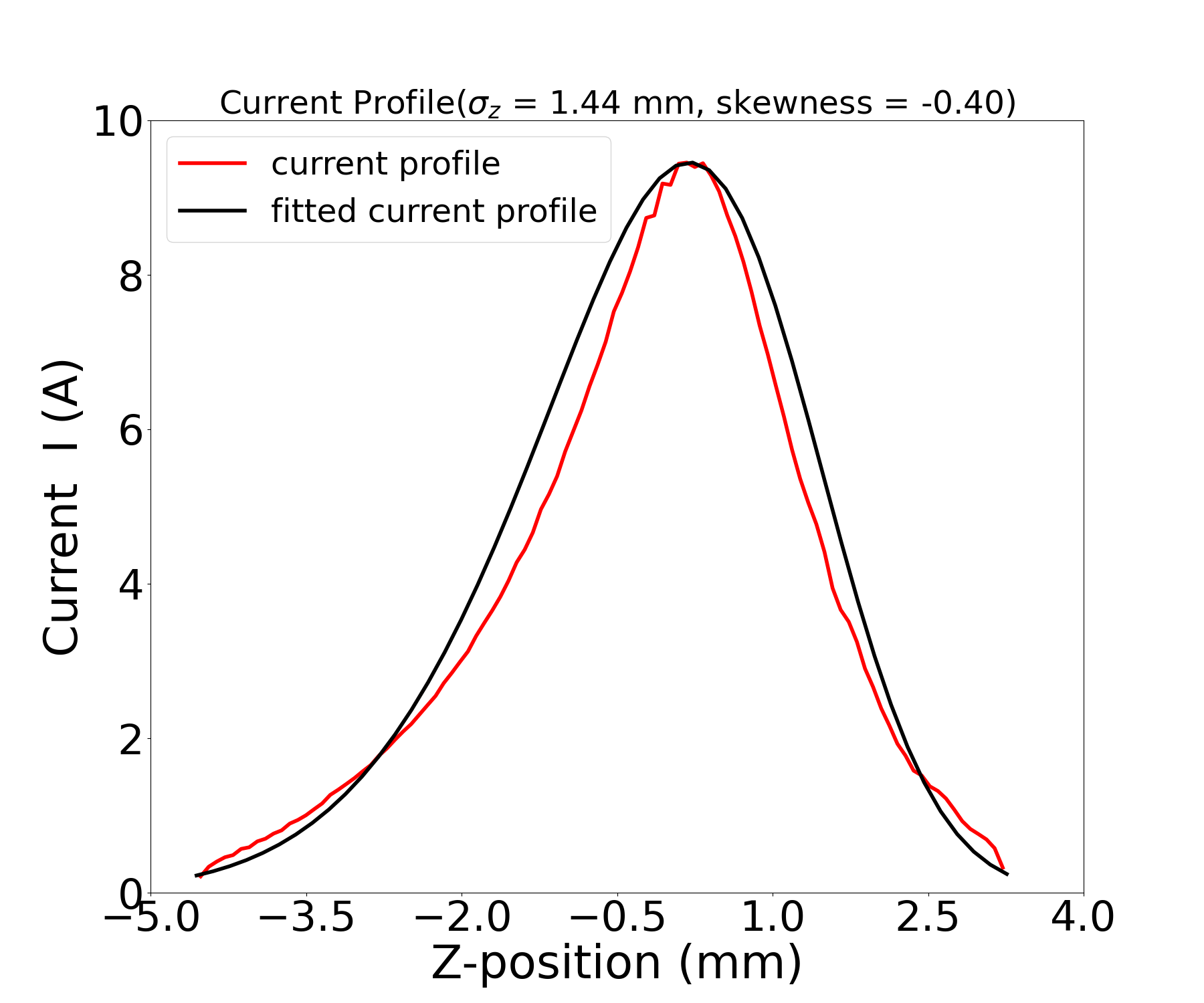}}
	\caption[width=1\textwidth]{The current profile of three cases with positive skewness(a), nearly zero skewness(b) and negative skewness(c). The current profile is plotted with the red line and the fitted current profile with skew-normal distribution is plotted with the black line. The head of the bunch is on the left of the coordinate.}
	\label{skew} 
\end{figure*}

In addition to the high-order energy spread, the nonlinear effects also result in the large asymmetry in the current profile at the exit of the injector, which is susceptible to the current-spike formation and leads to the current profile with a long-tail shape. The long-tail bunches are the potential to cause beam loss in the successive components along the beamline. Based on the shape of the current profile, the concept of skew-normal distribution is introduced to describe the profile. The skewness is the parameter of the distribution that reflects the degree of deviation from a normal distribution. The definition of skewness is:
\begin{equation}
	S = E[(\frac{X-\mu}{\sigma})^{3}]=\frac{\mu_3}{\sigma_3}=\frac{E[(X-\mu)^3]}{(E[(X-\mu)^2)])^{3/2}},
\end{equation}
where $X$ is the random distribution, $\mu$ is the mean, $\sigma$ is the standard deviation and $E$ is the expectation operator, $\mu_3$ is the third order central moment of the electron distribution. With this parameter introduced, the current profile can be fitted with a skew-normal distribution and the skewness parameter $S$ can reflect the approximate shape of the profile shape. The three typical current profile cases with different values of skewness are shown in Fig. \ref{skew}. In Fig. \ref{skew1}, the positive skewness reflects that the tail part of the bunch on the right side of the probability density function is longer than those on the left side and the majority of electron lies to the head half of the bunch. Instead, Fig. \ref{skew3} shows the distribution with negative skewness, which denotes the current profile whose majority tilts to the bunch tail. The slight asymmetry of these two kinds of the current profile is potential to be magnified by the strong density modulation in the downstream dispersive regions.

The longitudinal charge distribution $\rho_z$ is influential to the wakefield-induced energy modulation in beam delivery passing the accelerating cavity. The wake potential function is \cite{bane2016analytical}:
\begin{equation}
W(z) = -\int_{z}^{\infty} \rho_z(z^{\prime})w(z-z^{\prime}) dz^{\prime},
\end{equation}
where $w(z)$ is the Green's function of longitudinal wake potential of a point particle, and $\rho_z$ indicates the density of the longitudinal electron distribution along with the longitudinal position $z$. When the skewness value comes to almost zero, it signifies the current profile is nearly symmetric, which results in the more linear longitudinal energy modulation induced from the wakefield effect. Hence, the absolute value of the calculated skewness is treated as the fourth objective in the optimization.

\begin{figure}[htb] 
	\centering 
	\subfigure{\includegraphics*[width=0.8\linewidth]{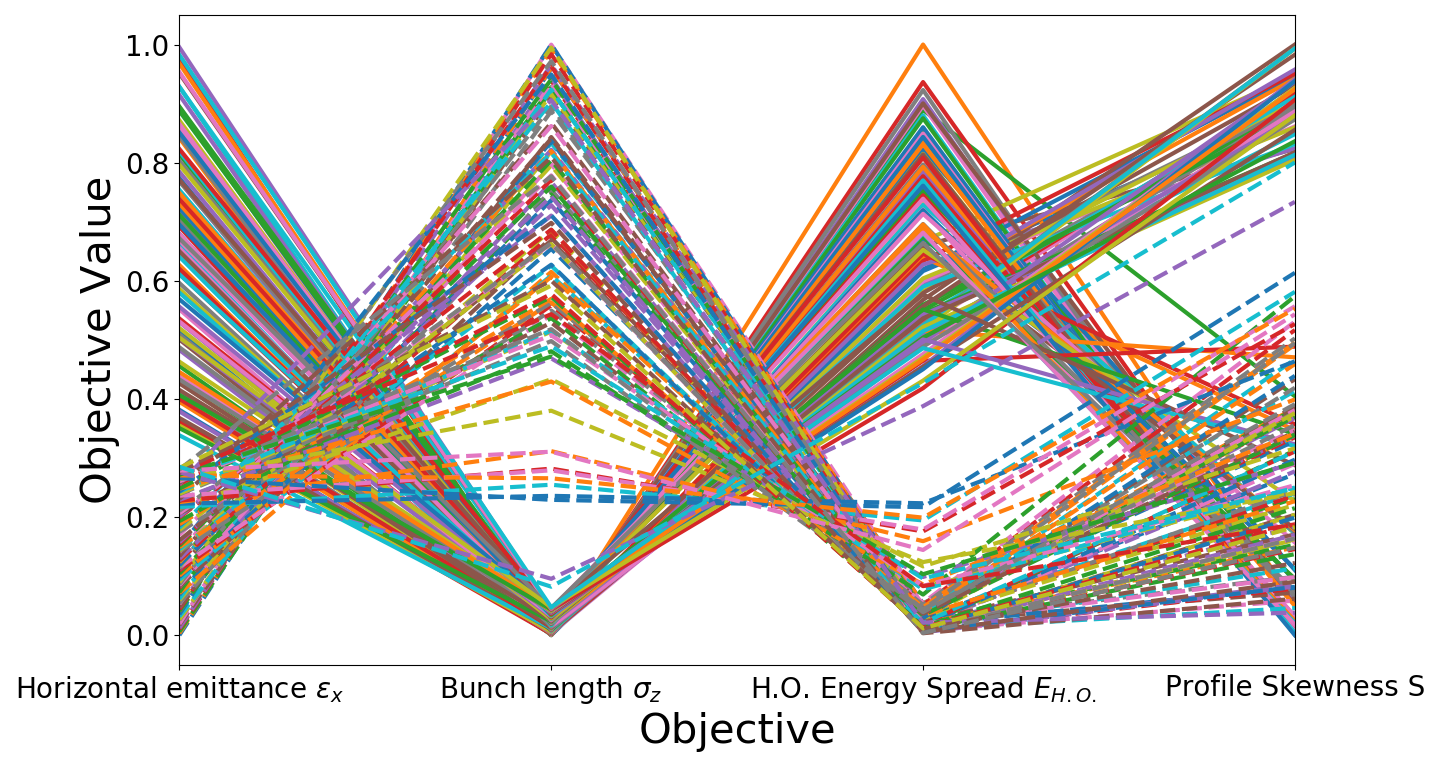}}
	\caption[width=1\textwidth]{The parallel coordinate plots of the 200 solutions in the final population of generation showing the objectives value path. 100 of them with the lowest transverse emittance (dashed line) and the other 100 lines present the solutions with the shortest bunch length (solid line). The four ticks in the objective value coordinate refer to the four objectives in the optimization.}
	\label{valuepath} 
\end{figure}

\begin{figure}[!htb] 
	\centering 
	\subfigure[]{
		\label{project1}	
		\includegraphics[width=0.38\linewidth]{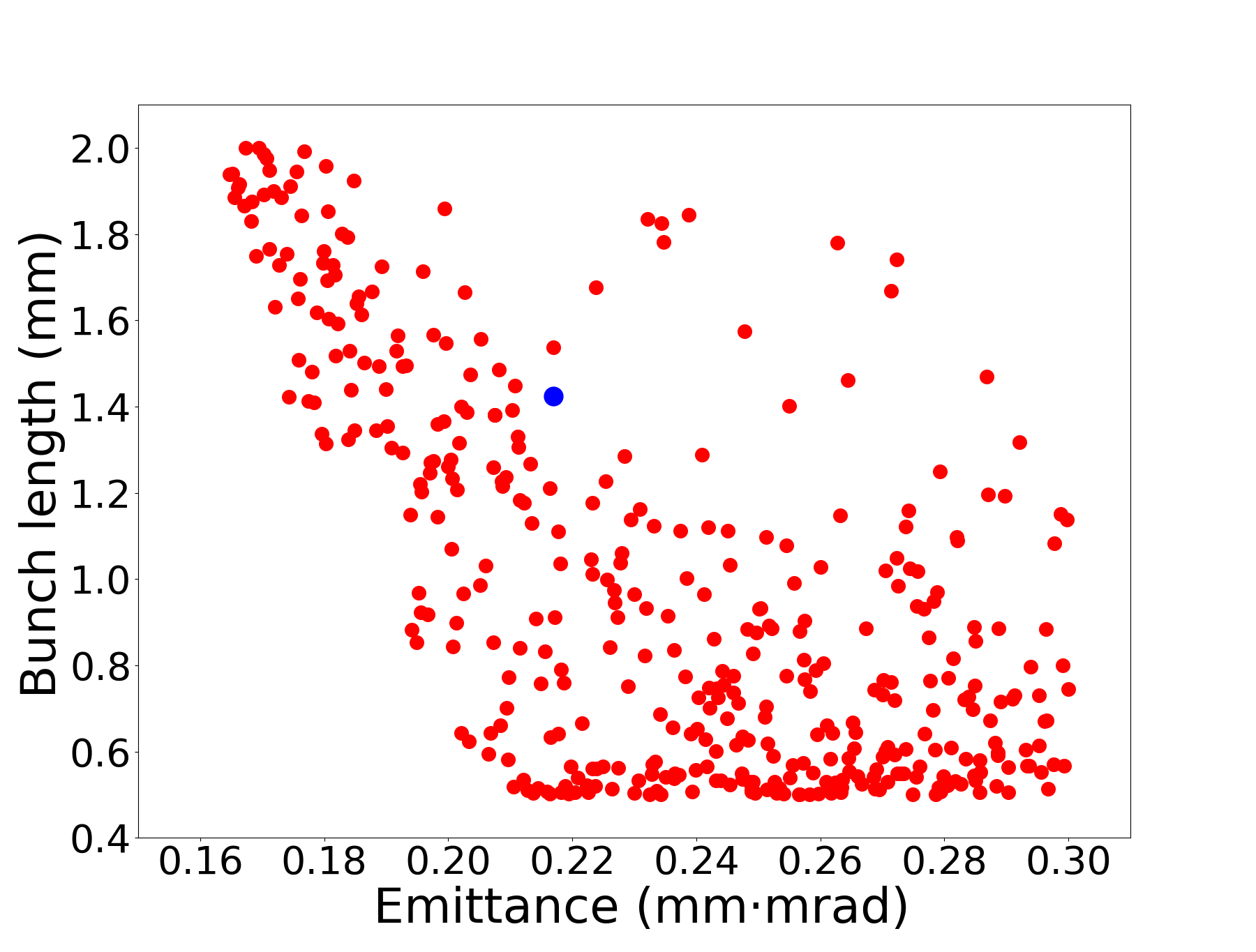}}
	\subfigure[]{
		\label{project2}	
		\includegraphics[width=0.38\linewidth]{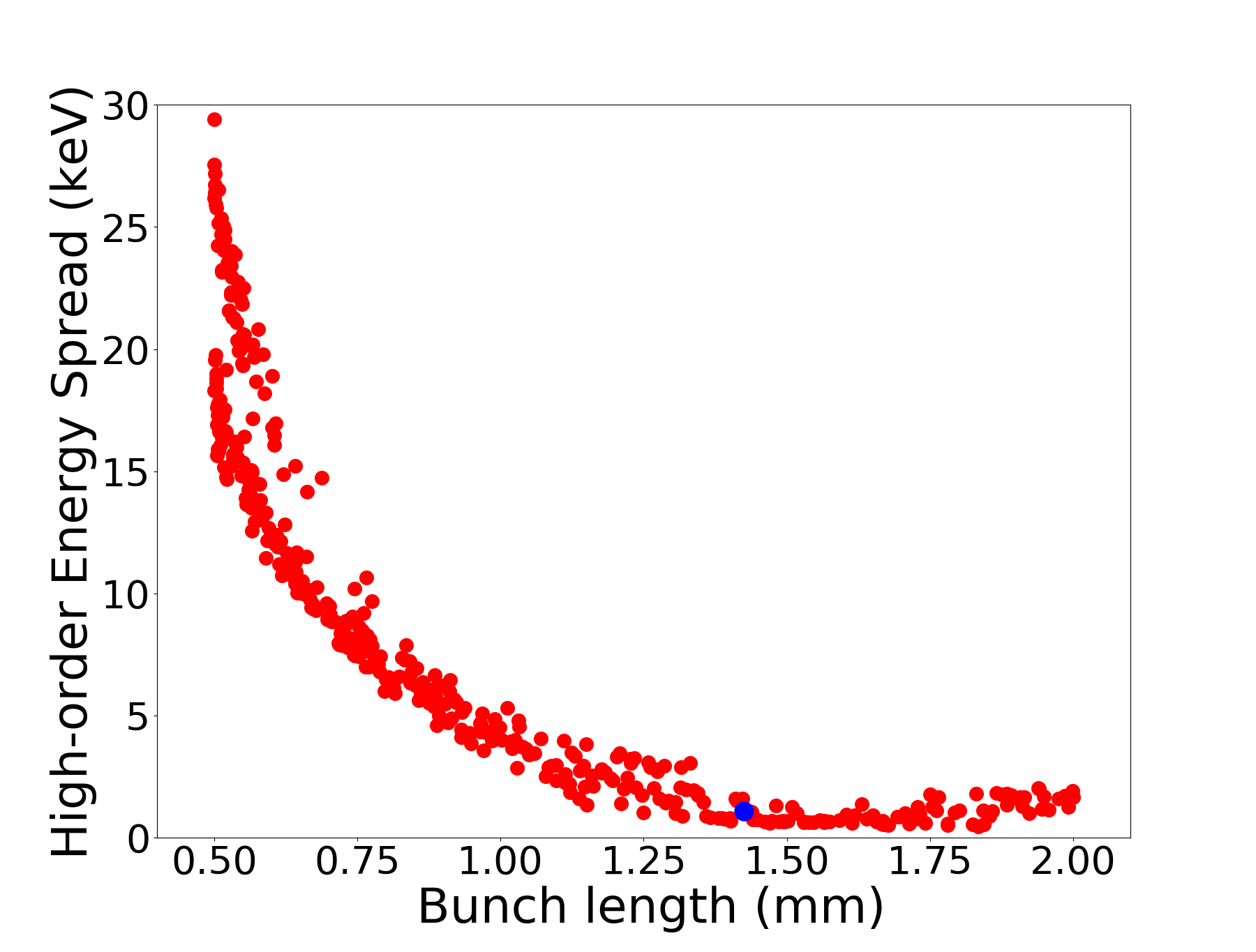}}
	\subfigure[]{
		\label{project3}	
		\includegraphics[width=0.38\linewidth]{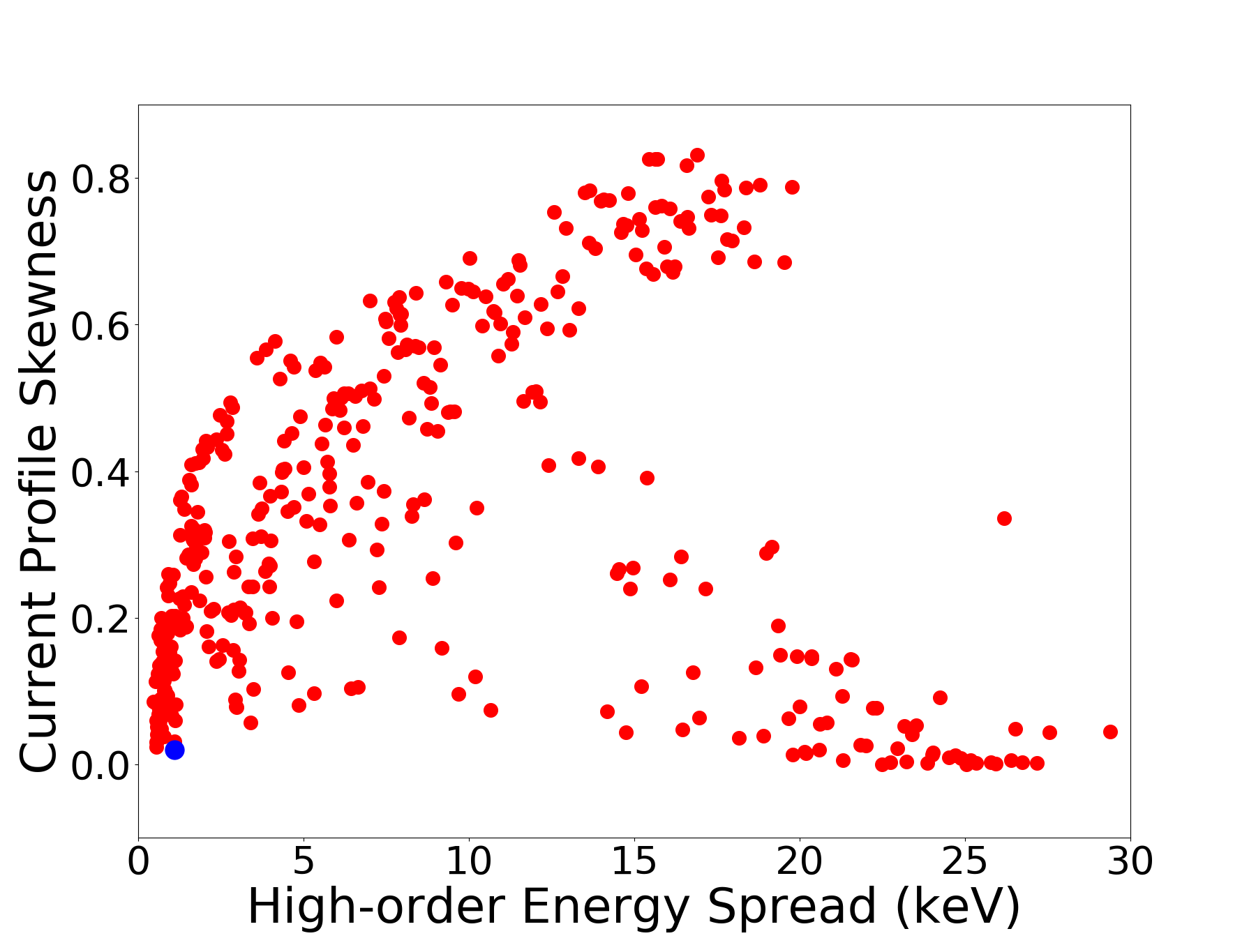}}
	\subfigure[]{
		\label{project4}	
		\includegraphics[width=0.38\linewidth]{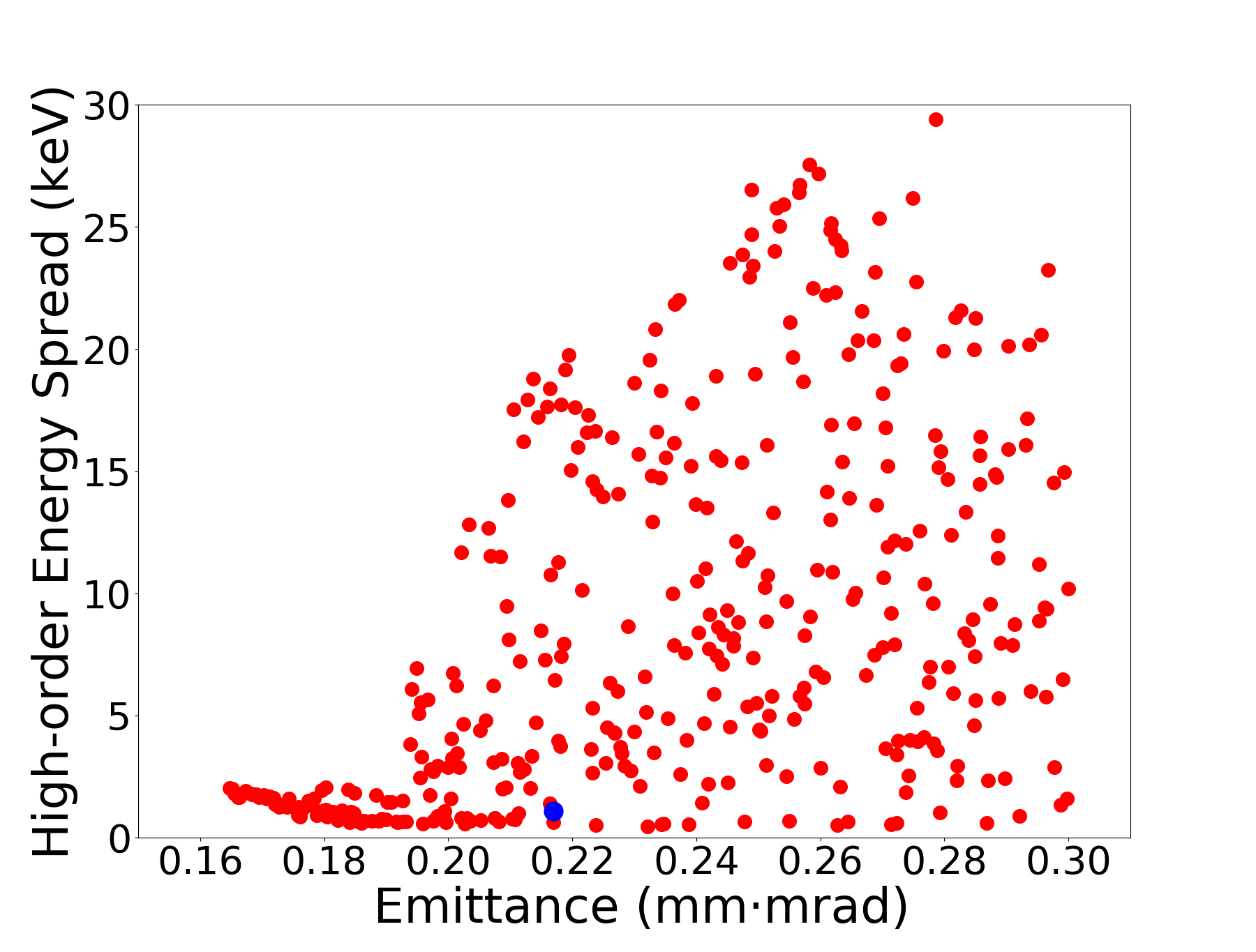}}
	\caption[width=1\textwidth]{The project of the Pareto front obtained from the last generation. The blue dots in each figure represent the selected solution of which the detailed beam dynamics parameters are shown in Fig. \ref{final_inj}.}
	\label{project} 
\end{figure}
The optimization result is presented in Fig. \ref{valuepath}, which shows 200 value-path plots for the objectives in the final result. The solutions are all derived from the population of the last generation in the optimization, 100 of them are the solutions with the lowest transverse emittance, the other 100 ones with the shortest bunch length. The values of each solution are normalized to the value between 0 and 1 by the following function:
\begin{equation}
	X_{nij}=\frac{X_{ij}-min(X_{i})}{max(X_{i})-min(X_{i})},
\end{equation}
where $X_{nij}$ is the normalized value, $X_{ij}$ is the $i$th objective of the $j$th solution, $max(X_{i})$ is the maximum value of the $i$th objective among the whole population, $min(X_{i})$ is the minimum value of the $i$th objective among the solutions. It is difficult to present the visible correlation between the objective parameters in the Pareto-optimal front of the optimization result because of the four-dimension Pareto-Optimal front. Some strong correlations between the optimization objectives can be observed and the potential relevance can be figured out in the projections of the front which is presented in Fig. \ref{project}.

The two beam dynamics properties, transverse emittance and bunch length, are usually selected as the two objectives in the injector beam dynamics optimization strategy \cite{gulongitudinal,sannibale2019high,schmerge2014lcls,papadopoulos2012injector,bazarov2009maximum,filippetto2014maximum}. Normally, the tradeoff between the two parameters is obvious in particle accelerator beam dynamics design. In this task, however, as shown in Fig. \ref{project1}, the correlation is ambiguous and this weak negative relevance indicates that a relatively longer bunch is requisite for achieving the needed ultra-low transverse emittance. This is totally due to the other two involved objective parameters which are influential to the final multi-dimension Pareto front distribution. More detailed parameters rather than bunch length are explored and some potential correlation can be found in Fig. \ref{project2} and Fig. \ref{project3}. Regarded as one of the main factors to the nonlinear bunch longitudinal compression, the high-order energy spread is largely induced from the strong nonlinear space charge effect inside the bunch and RF curvature in the cavities, especially nonuniform energy modulation in the buncher cavity during the low $\beta$ transporting range. It can be partially mitigated through lengthening the bunch length, which is shown in Fig. \ref{project2}. This clear correlation provides helpful and effective guidance for current-horn suppression in strong magnetic longitudinal compression. As for the current profile skewness, the longitudinal density distribution along the bunch is largely modulated by the harmonic RF mode in the buncher of which the voltage and phase of the mode play the crucial roles in shaping the desired current profile after the velocity compression, which can be seen in Fig. \ref{skew}. Fig. \ref{project3} indicates the nearly symmetric current profile can be achieved with a small high-order energy spread if the bunch length is relatively longer. Additionally, the results present some solutions with relatively low emittance and low high-order energy correlation, which can be verified in Fig. \ref{project4}. 

\begin{figure*}[!htb] 
	\centering 
	\subfigure[]{
		\label{1emittance}	
		\includegraphics[width=0.3\linewidth]{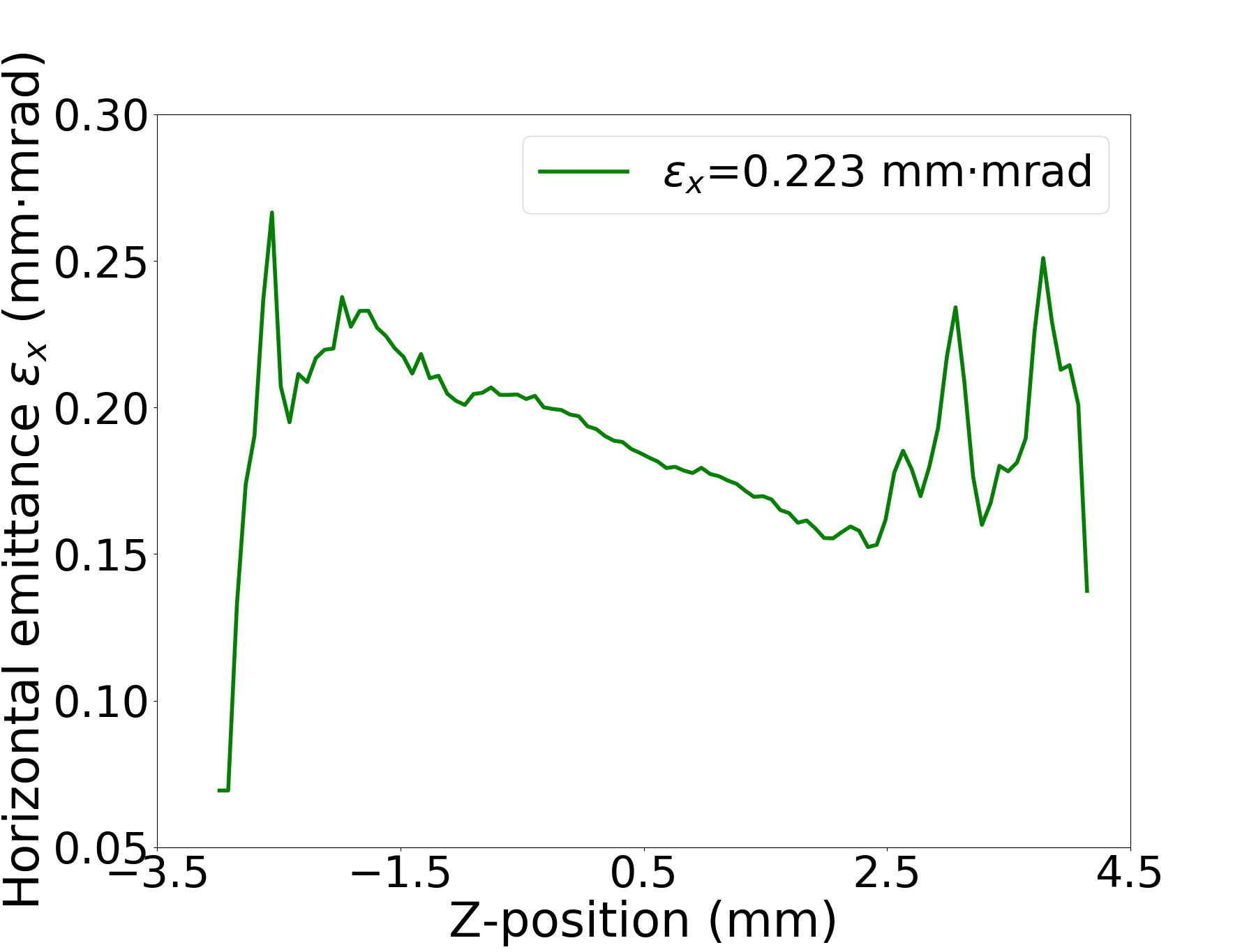}}
	\subfigure[]{
		\label{1profile}	
		\includegraphics[width=0.3\linewidth]{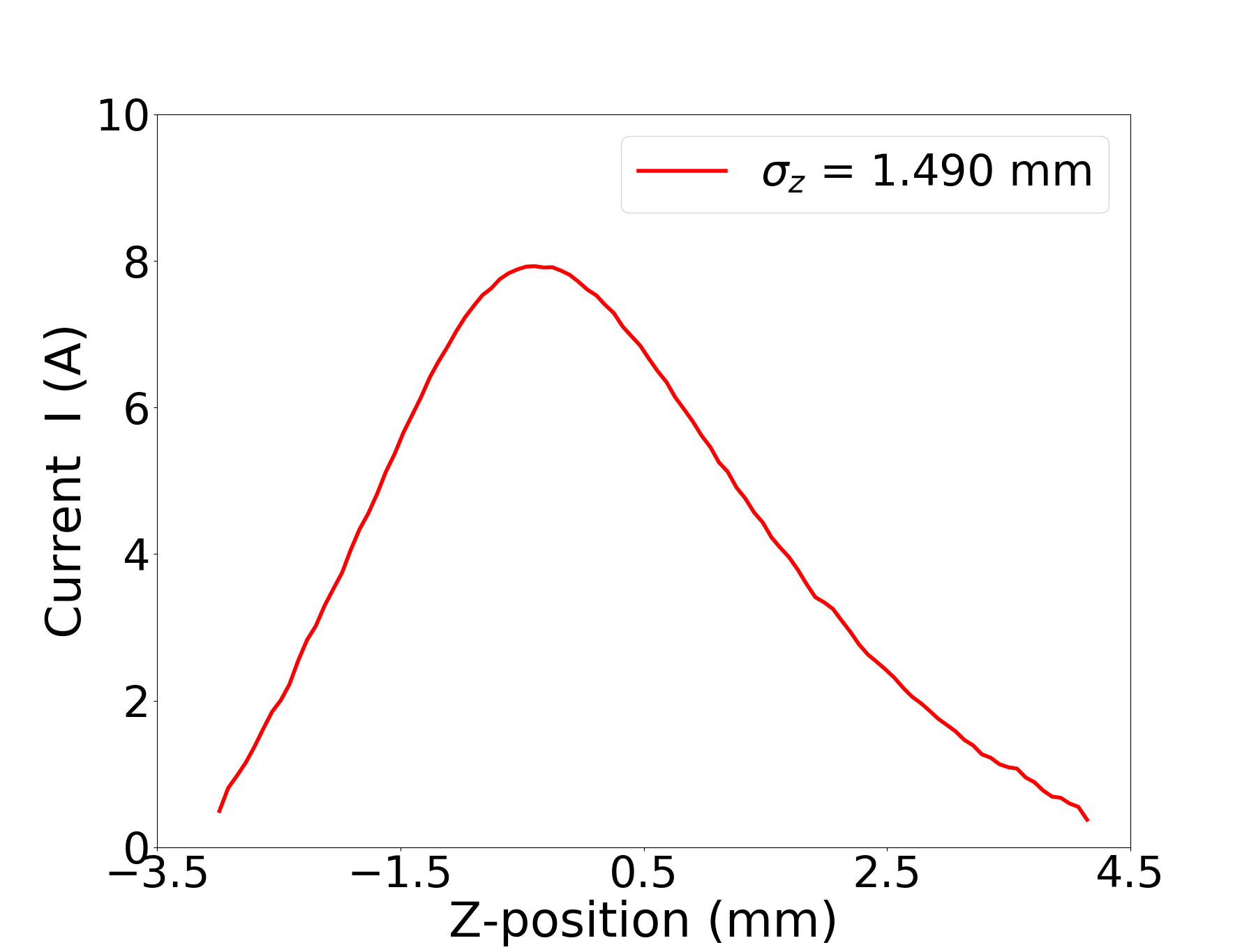}}
	\subfigure[]{
		\label{1spread}	
		\includegraphics[width=0.3\linewidth]{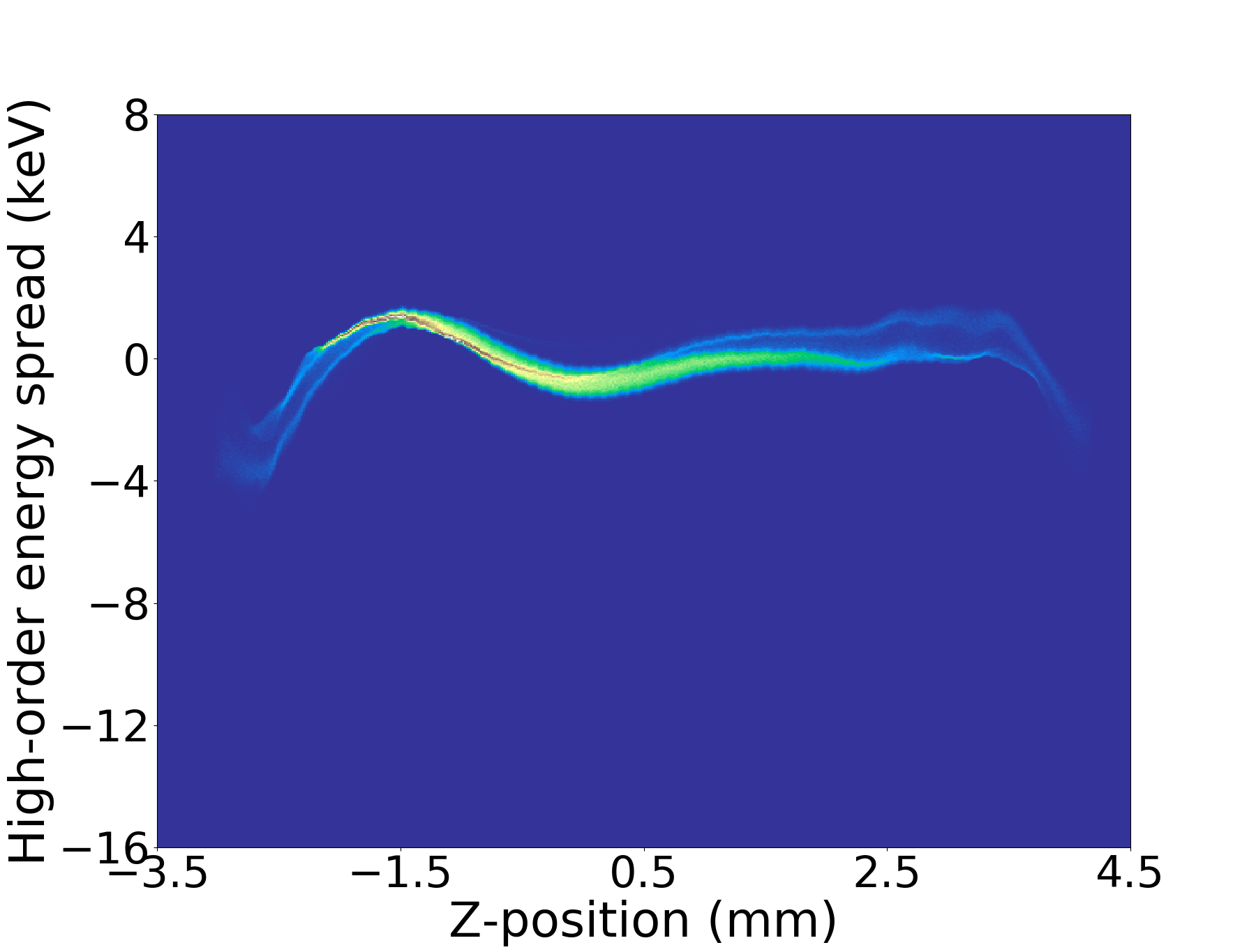}}
	\subfigure[]{
		\label{2emittance}	
		\includegraphics[width=0.3\linewidth]{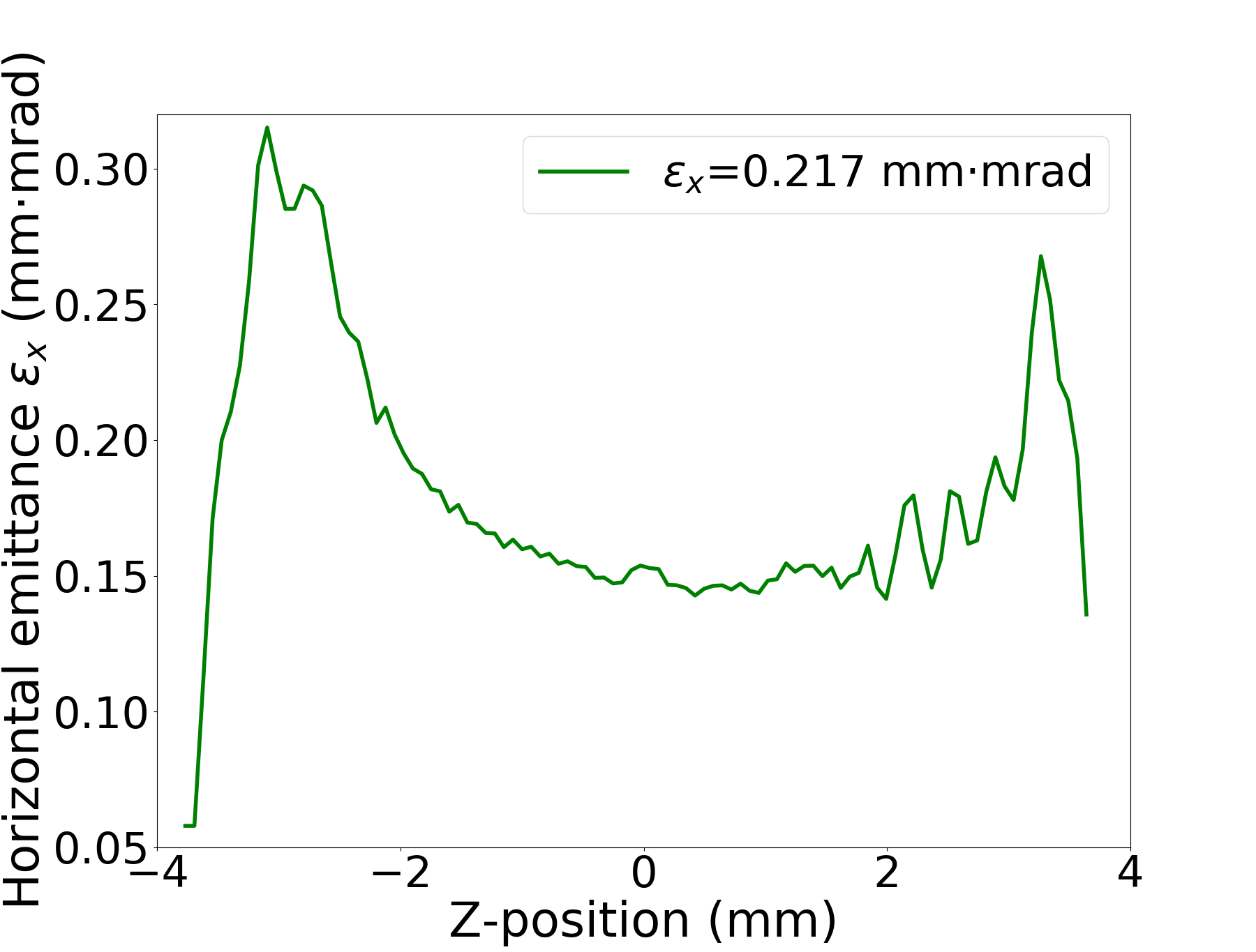}}
	\subfigure[]{
		\label{2profile}	
		\includegraphics[width=0.3\linewidth]{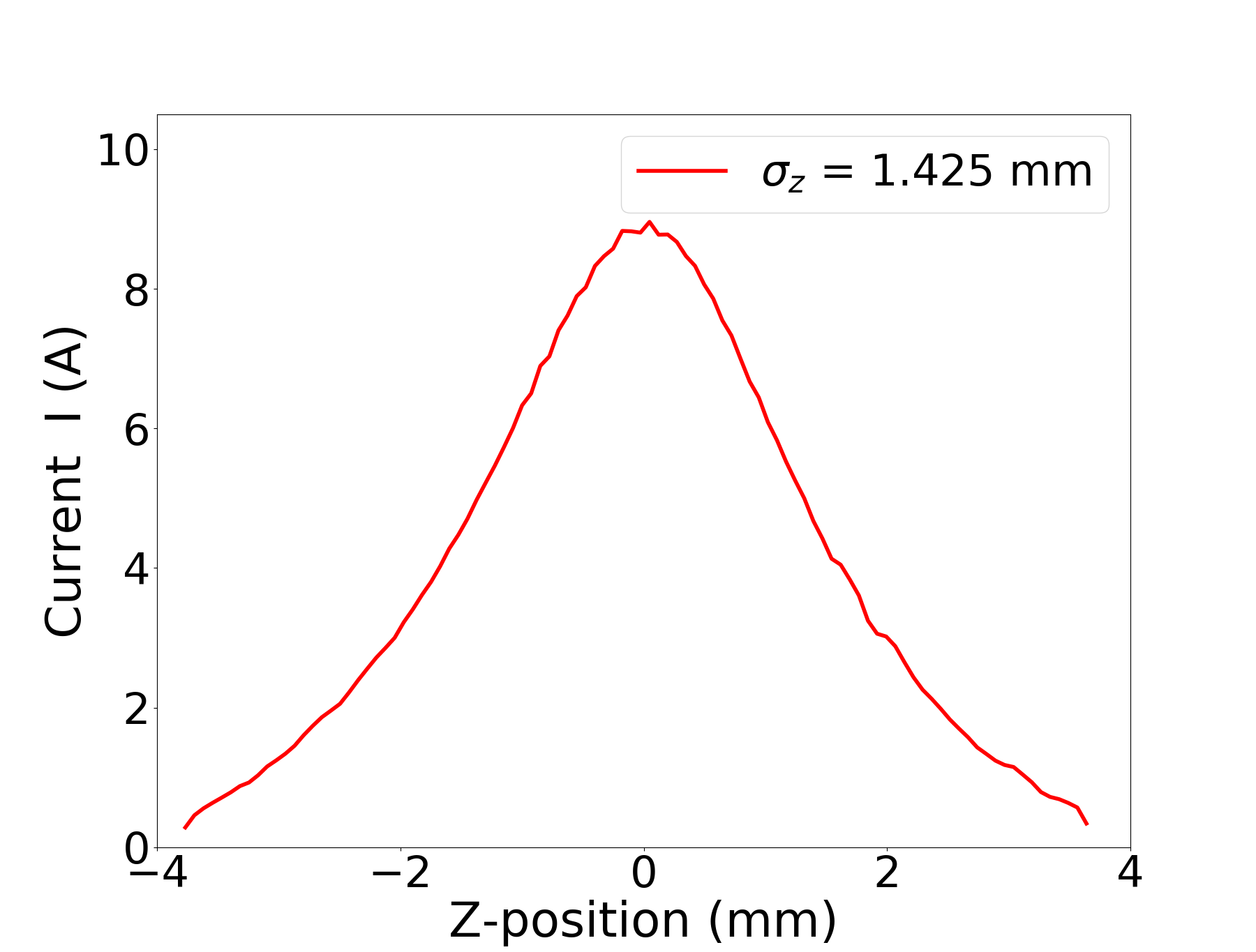}}
	\subfigure[]{
		\label{2spread}	
		\includegraphics[width=0.3\linewidth]{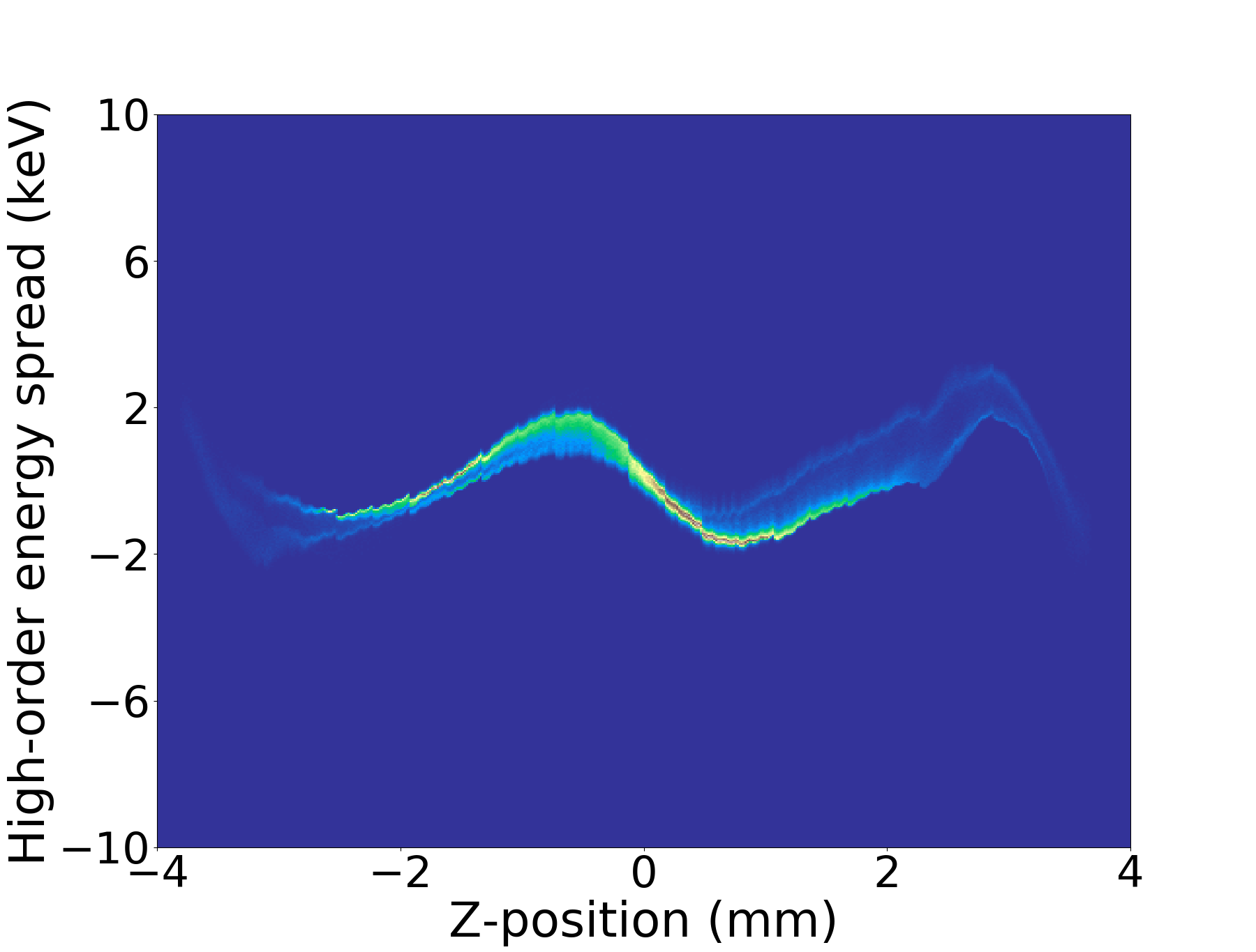}}
	\caption[width=1\textwidth]{Beam dynamics result in comparison between the initial physical design (top row) and optimized design (bottom row) with longitudinal phase space compensation technology applied. The left side of the Z coordinate represents the head of the bunch. The longitudinal phase space distribution difference at the injector exit will immensely affect the final current profile downstream of the multi-stage magnetic chicanes.}
	\label{final_inj} 
\end{figure*}

However, as a tradeoff, the long bunch length will result in stronger magnetic compression required in the multi-stage chicanes to realize the desired peak current. The intense compression coefficient $R_{56}$ may cause an unwanted CSR effect and current spike formation, both of which degrade the FEL performance. In SHINE FEL beamline design, the requirement of the electron bunch longitudinal phase space for different FEL lasing modes differs from each other, thus a tradeoff between the three objectives other than transverse emittance should be balanced elaboratively. Consequently, the transverse emittance of the selected solution is 0.217 mm·mrad with a bunch length of 1.425 mm, and the skewness of the current profile is 0.02 with the high-order correlated energy chirp is 1.09 keV.

Based on the optimization result shown above, the detailed beam dynamics simulation result of the selected optimal setting points is shown in Fig. \ref{final_inj} under the comparison with the initial physical design, which is mainly aimed at eliminating the third and above terms of energy spread \cite{zhu2021many}. With longitudinal phase space compensation by means of the dual-mode buncher, the most distinct difference comes to the longitudinal distribution inside the bunch. Fig. \ref{1profile} shows the original electron distribution of which the peak current is skewed to the head of the bunch. While the current profile shape after the compensation is demonstrated in Fig. \ref{2profile}, whose profile is corrected to be more symmetric in the beam longitudinal distribution, which is preferable for the longitudinal compression in the downstream magnetic chicanes. In addition, Fig. \ref{2emittance} presents the transverse projected emittance is slightly lower with the slice emittance less than 0.2 mm·mrad in the core of the beam. This ultra-low slice emittance facilitates the transverse matching with the machine lattice in the undulator, which is vital for FEL lasing performance. Also, the bunch length property is shorter slightly in consideration of the balance between the achievable compression strength and maintenance of the residual high-term correlated energy chirp. Additionally, as shown in Fig. \ref{2spread}, with the proposed dual-mode buncher applied, the curvature in the high-order energy spread can be optimized to a more symmetric and reverse shape compared with the original design which is shown in Fig \ref{1spread}. This improvement facilitates compensation of the nonlinear compression and prevention of the current spikes.

\section{start-to-end optimizations}
\label{s3}

\subsection{Longitudinal dynamics optimization for bunch compression}
\label{s31}
\begin{figure*}[htb]
	\centering
	\includegraphics[width=0.95\linewidth]{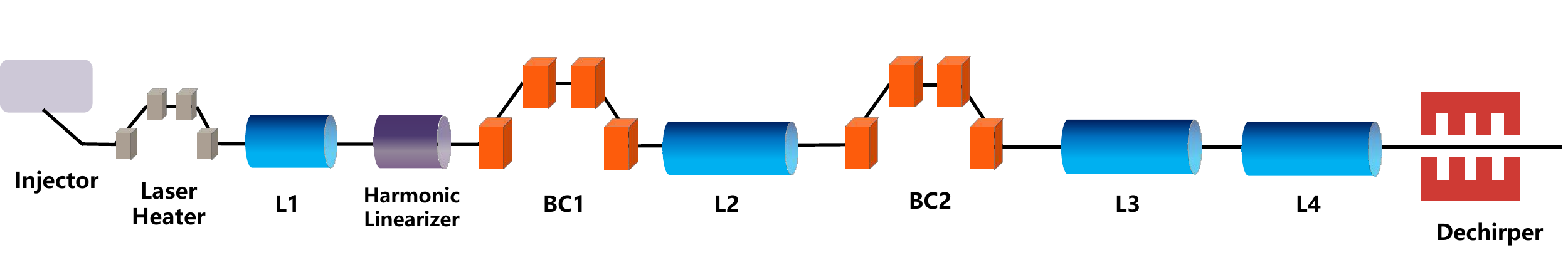}
	\caption{The conceptual schematic of the SHINE linac section, accelerating the beam to 8 GeV with a peak current of 1.5 kA. The 3rd harmonic cavity is deployed as the linearizer of bunch longitudinal phase space. The two sets of bunch compressor chicane (BC1 and BC2) are both 4-dipole chicane which converts the energy distribution in the bunch into density modulation. The corrugated structure is employed to provide a wakefield-induced energy compensation to linearize the energy chirp in the bunch.}
	\label{linac}
\end{figure*} 
	
In the last section, the detailed approach and optimization in the photoinjector are described and demonstrated. The final goal of the optimization is improving the longitudinal phase space distribution after the multi-stage bunch compression at the exit of the linac. Considering the balance between the accuracy and time efficiency, the one-dimensional longitudinal tracking code LiTrack \cite{bane2005litrack} is introduced in this section. The main optimization objective is the shape of the current profile which is preferred to be a flat-top shape. The optimization result is verified by the ELEGANT \cite{borland2000elegant} simulation program. The layout of the main accelerating section is shown in Fig. \ref{linac}.

The relative energy deviation $\delta$, correlates with the position Zi, and it leads to the differences in transmission path length between the electrons in the process of longitudinal density modulation. The definition of the relative energy deviation $\delta$ is:
\begin{equation}
	\delta=\delta_0+c_1z_i+c_2{z_i}^2+c_3{z_i}^3+O({z_i}^4),
	\label{eq1}
\end{equation}
where $\delta_0$ is the initial uncorrelated energy spread, $c_1$, $c_2$ and $c_3$ are the first-, second-, and third-order coefficient of the energy chirp, respectively. The magnetic chicane compressor is usually approximately defined by its first-, second-, and third-order dispersion coefficient values of $R_{56}$, $T_{566}$, and $U_{5666}$. In the 4-dipole chicane, the final longitudinal position $z_f$ of the electron is given by:
\begin{equation}
	z_f=z_i+R_{56}\delta+T_{566}\delta^2+U_{5666}\delta^3+O({\delta}^4),
	\label{BC}
\end{equation}
\begin{equation}
	R_{56}\approx-\theta^2(2D+\frac{4}{3}L_b),
\end{equation}
\begin{equation}
	T_{566}\approx-\frac{3}{2}R_{56},
\end{equation}
\begin{equation}
	U_{5666}\approx2R_{56},
\end{equation}
where the $z_i$ is the initial position of the particle, $\theta$ is the bending angle of dipoles in the chicane, $D$ is the longitudinal distance between the first and second dipoles (also the third and fourth ones), $L_b$ is the length of each rectangular dipole. 

Substituting the Eq. \ref{eq1} into the Eq. \ref{BC} gives: 
\begin{equation}
    \begin{aligned}
	z_f&=(1+c_1R_{56})z_i+(c_2R_{56}+c_1^2T_{566})z_i^2\\
	&+(c_3R_{56}+2c_1c_2T_{566}+c_1^3U_{5666})z_i^3+O(z_i^4).
	\label{BCfinal}
	\end{aligned}
\end{equation}
This analytical expression indicates that the nonlinear (especially third-order term $c_3$) energy chirp in the bunch longitudinal phase space is not the only cause for the nonlinear compression, for a normally-used 4-dipole C-shape chicane, even the linear chirp (the first-order coefficient $c_1$) can result in the high-order effect after being squared and cubed in the Eq. \ref{BC}. Actually, the detailed term values of the longitudinal phase space and the bunch compressor chicane are highly interacted with each other, which results in the nonlinear deformations of the phase space after the compression, together with the formation of spike in the current profile.

Here, the multi-objective optimization algorithm NSGA-II is introduced together with the one-dimensional longitudinal beam dynamics tracking code to conduct the optimization. The numbers of generation and population within each generation are set to 150 and 200. The final bunch length is chosen as one optimization objective. Another objective comes to the current profile characteristic value which is defined to roughly describe the current profile shape. This objective parameter is obtained by evaluating two values. The first parameter is named the ``shape parameter", which is introduced to describe the basic profile shape after LiTrack simulation. The other one is set to describe the portion of the bunch core in the profile. The bunch core refers to the region where the current is higher than $90\%$ of the peak current. Hence, the characteristic value is defined as:
\begin{equation}
	k=1-k_1-k_2=1-\frac{S_{tp}}{S}-0.5\times\frac{H_{pr}}{H},
	\label{cp}
\end{equation}
where $k_1$ and $k_2$ refer to the ``shape parameter" and ``core parameter", respectively. $S$ is the area enclosed by the current profile, $S_{tp}$ is the area of the top half of the current profile, which is divided by the line that refers to the half value of the peak current. $H$ is the number of bins in the histogram computation for the current profile. $H_{pr}$ is the number of bins in the bunch core. The coefficient $0.5$ is multiplied to scale the two values to the same variable range. Therefore, the smaller the current profile characteristic value, the flatter and wider the current profile will be, which is desirable and preferable for driving the FEL radiation in the undulators downstream the beamline.

\begin{figure}[htb] 
	\centering 
	\subfigure[]{
		\label{profile1}
		\includegraphics[width=0.3\linewidth]{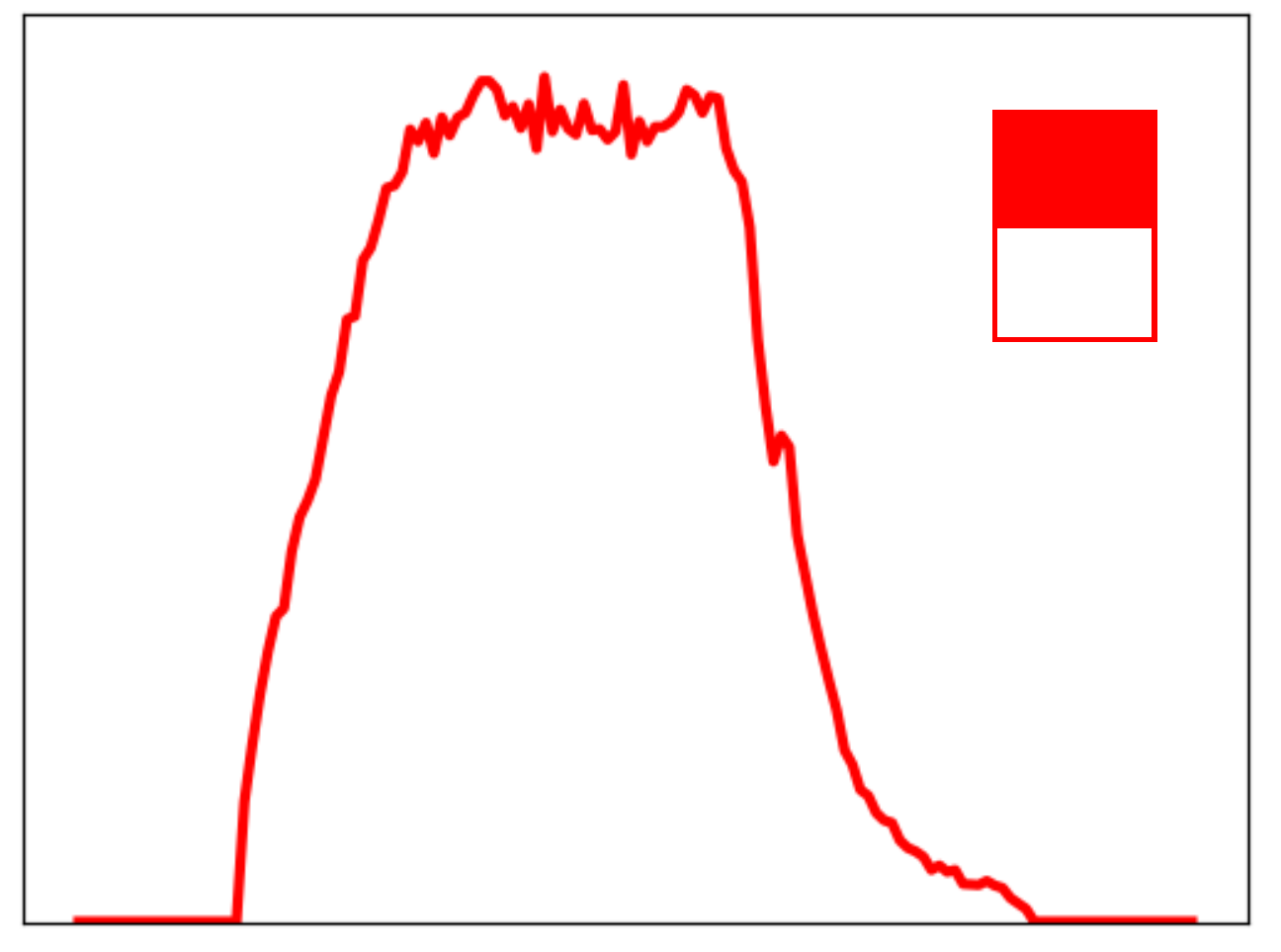}}
	\subfigure[]{
		\label{profile2}
		\includegraphics[width=0.3\linewidth]{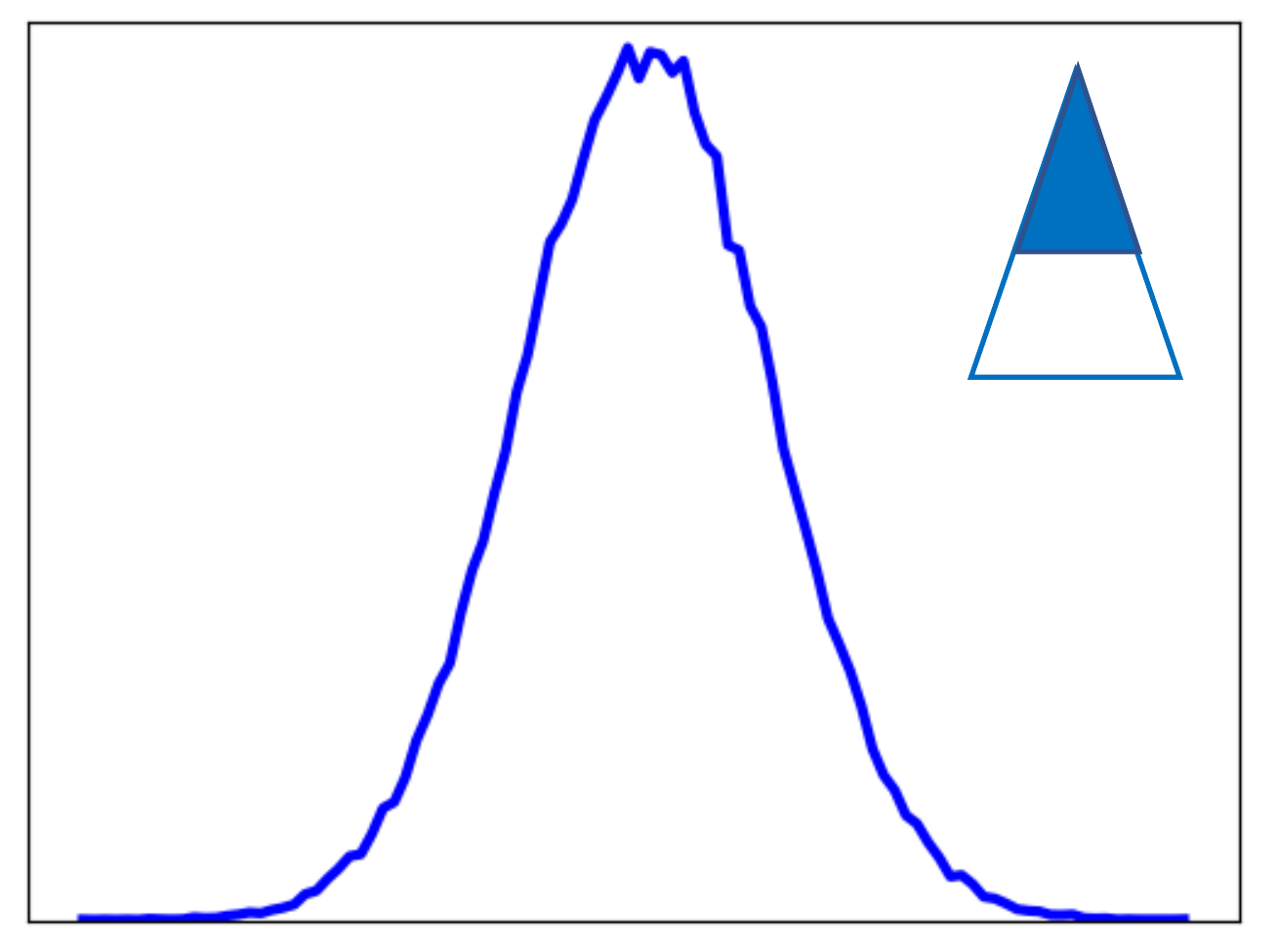}}
	\subfigure[]{	
		\label{profile3}
		\includegraphics[width=0.3\linewidth]{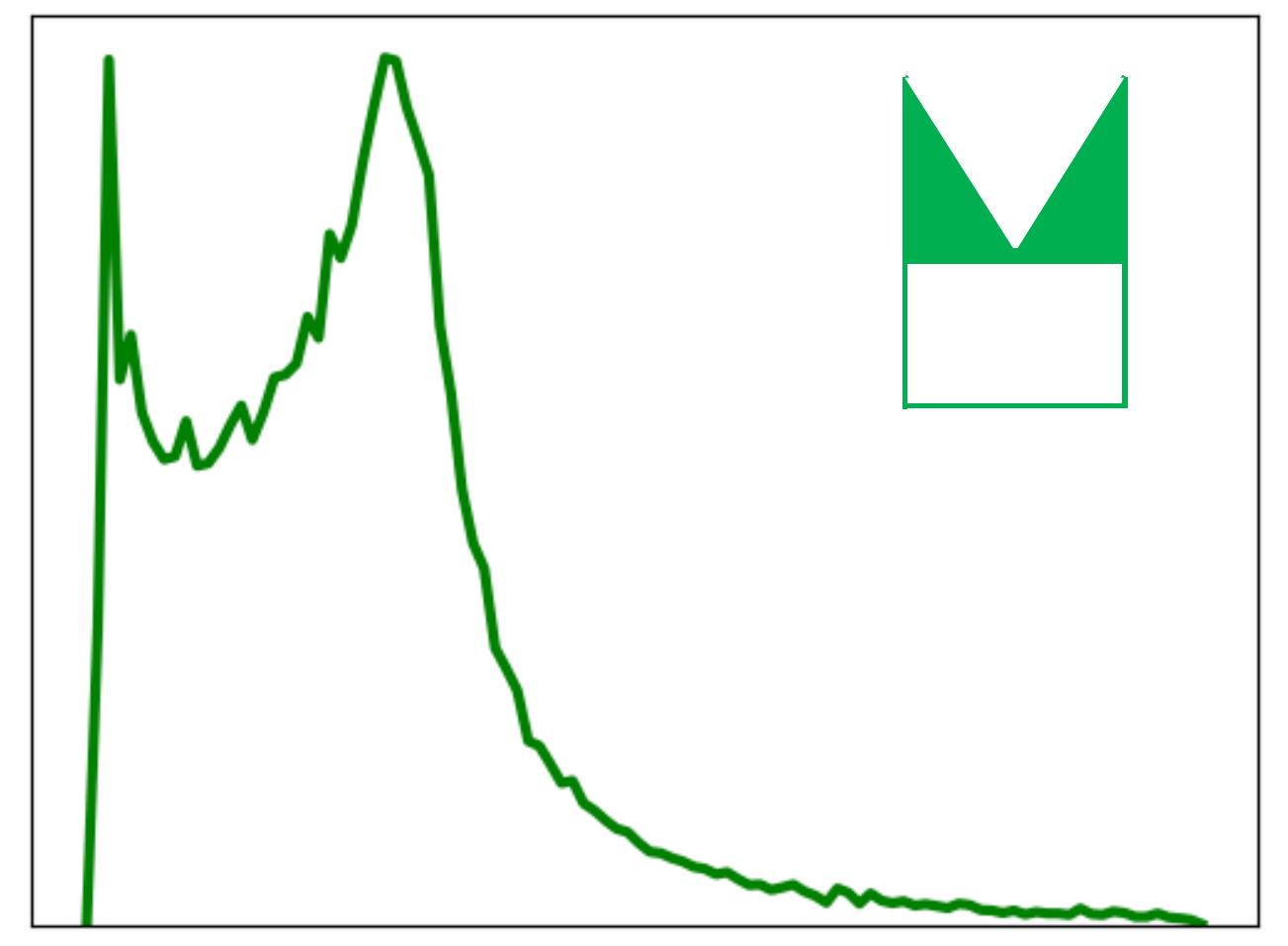}}
	\caption[width=1\textwidth]{The three different kinds of current profile shape after bunch compression. From left to right is the shape of ``flat-top" (a), ``gaussian-like" (b) and ``double-horn" (c).}
	\label{profiles} 
\end{figure}

In Fig. \ref{profiles}, three typical scenarios of the current profile are presented. Fig. \ref{profile1} shows the uniform longitudinal density distribution which is desirable for better control of the FEL performance. Fig. \ref{profile2} shows the Gaussian distribution profile, which can be achieved with the lower nonlinear energy spread, i.e. small high-order terms $c_3$ in a polynomial fit of correlated energy spread $\delta$. This kind of profile shape is not preferable for its nonuniform, which can not facilitate stability of FEL lasing performance. Moreover, Fig. \ref{profile3} shows the typical double spikes that often appear due to the nonlinearity in the strong bunch compression. Therefore, the flat-top distribution is the most desirable profile shape for the SHINE undulator section, especially the externally seeded harmonic cascade FELs as it facilitates the overlapping between the radiation pulse from the first radiator with the electron bunch in the second cascading stage.

\begin{figure}[htb] 
	\centering 
	\includegraphics[width=0.6\linewidth]{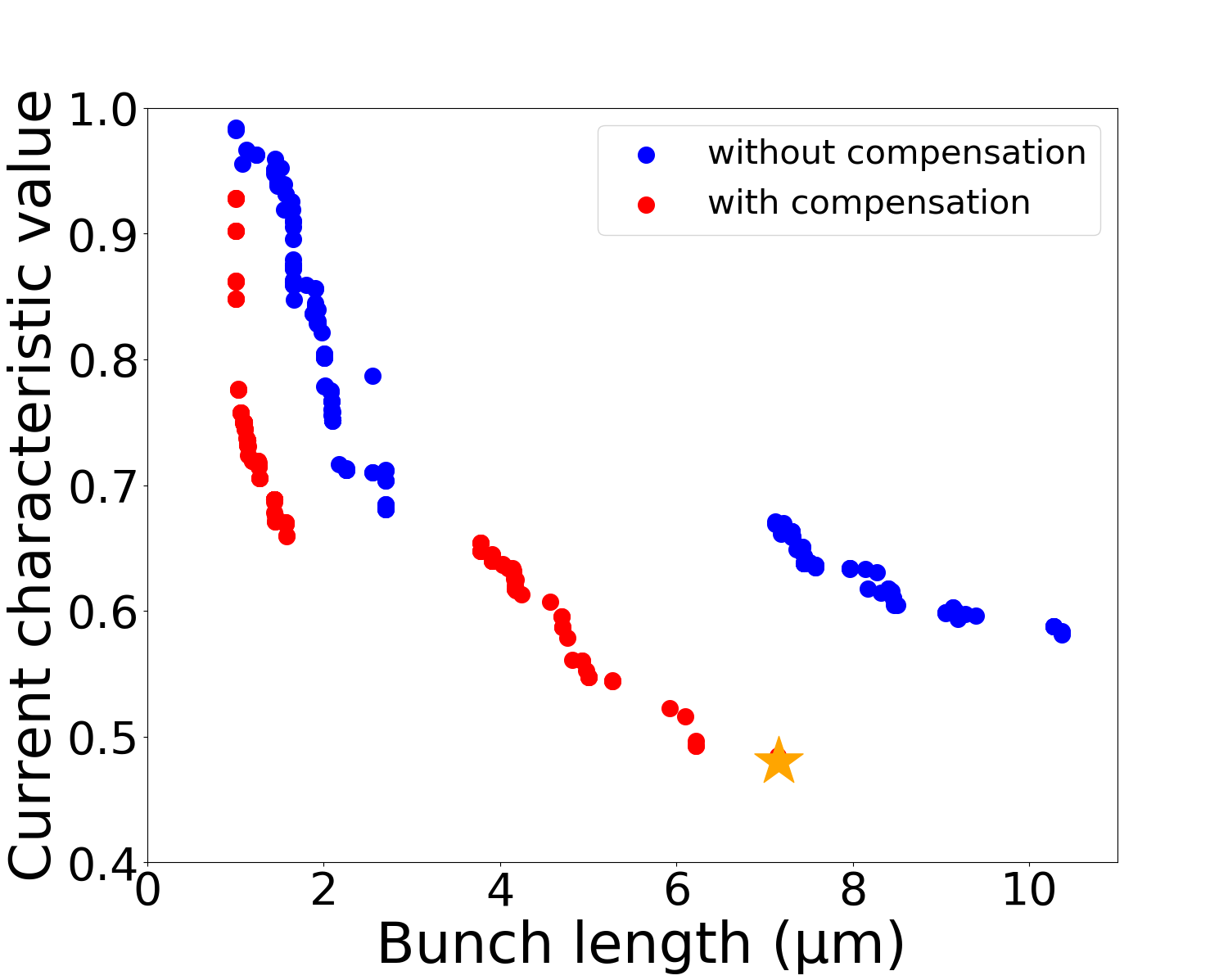}
	\caption[width=1\textwidth]{The comparison of Pareto front after 150 generations between the two different input beam longitudinal phase space distribution. The blue dots group presents the solutions from previous design optimization and the red ones are generated from the optimization with the compensation method, the orange star represents the selected optimal solution.}
	\label{comparison} 
\end{figure}

In this optimization strategy, the selected injector simulation result in the previous section was set as the input beam longitudinal information for the beam physics tracking and optimization in the linac. The Pareto front of the optimization is present in Fig. \ref{comparison}, together with the result of the one without longitudinal phase space compensation using the dual-mode buncher. It can be observed that it is such a tough task to generate the specific uniform bunch longitudinal density distribution with high peak current with bunch charge of 100 $\rm pC$, as the obvious tradeoff showed in Fig. \ref{comparison}. The transformation of the bunch longitudinal density results from the nonlinear effects in the chicane sections, the tiny perturbation of the energy chirp in the longitudinal phase space will make a destructive impact on the longitudinal distribution especially under the stronger compression scenario (shorter bunch length), and this often leads to the generation of the current spike (larger profile characteristic value).

\begin{figure}[htb] 
	\centering 
	\includegraphics[width=0.6\linewidth]{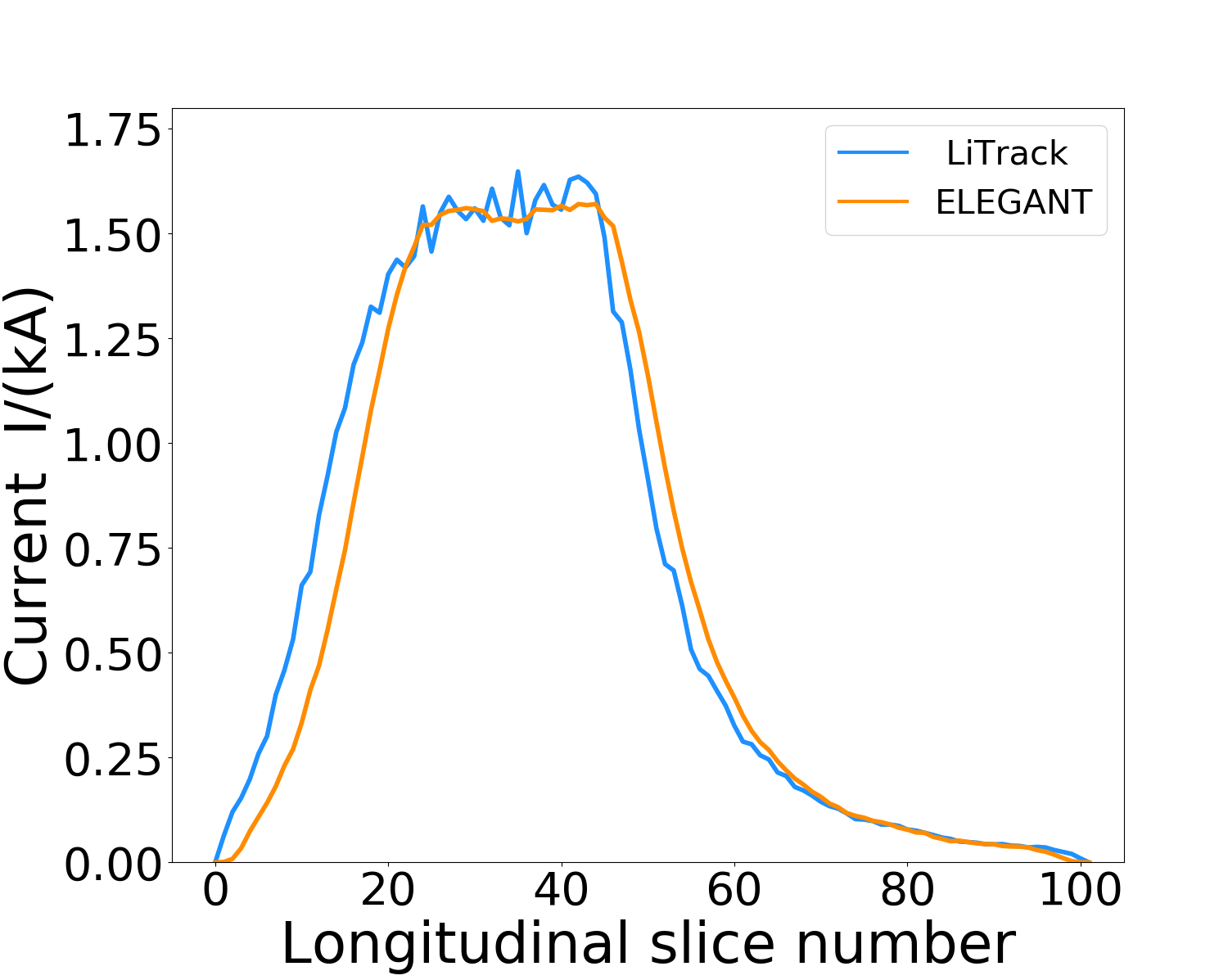}
	\caption[width=1\textwidth]{The current profile shape at the end of the linac section simulated by the LiTrack code (blue) and ELEGANT (orange).}
	\label{profile_comparison} 
	
\end{figure}

Based on the Pareto front from optimization, the selected solution comes to the one with an RMS bunch length of 7 $\rm \mu m$ whose longitudinal property exists a relatively uniform core in the distribution. Though the compression scenario is not too extreme to generate an ultra-short bunch with a peak current of over 2 $\rm kA$, it is advantageous to provide electron bunch with relative flat-top longitudinal distribution, which is one of the essential features for driving both the SASE beamline and externally seeded FEL beamline simultaneously. 

Since LiTrack is a fast longitudinal tracking code whose merit is the quick computational speed, its simplicity facilitates the fast beam dynamics longitudinal optimization. However, it does not involve collective effects such as the LSC effect and CSR effect which have an intricate energy modulation on beam longitudinal phase space quality. Therefore, the 6-dimensional simulation code ELEGANT is introduced to accurately track the beam delivery in the linac to verify the simulation result from LiTrack code. Here it should be mentioned that the bending  angle parameter in the second magnetic chicane is tweaked slightly (less than 2\% in R56 value) to match the bunch length after compression. In addition, the more accurate and detailed collective effects study will be conducted in the following work.

\begin{figure*}[htb]
	\centering
	\subfigure[]{
	    \label{lps1}
	    \includegraphics[width=0.3\linewidth]{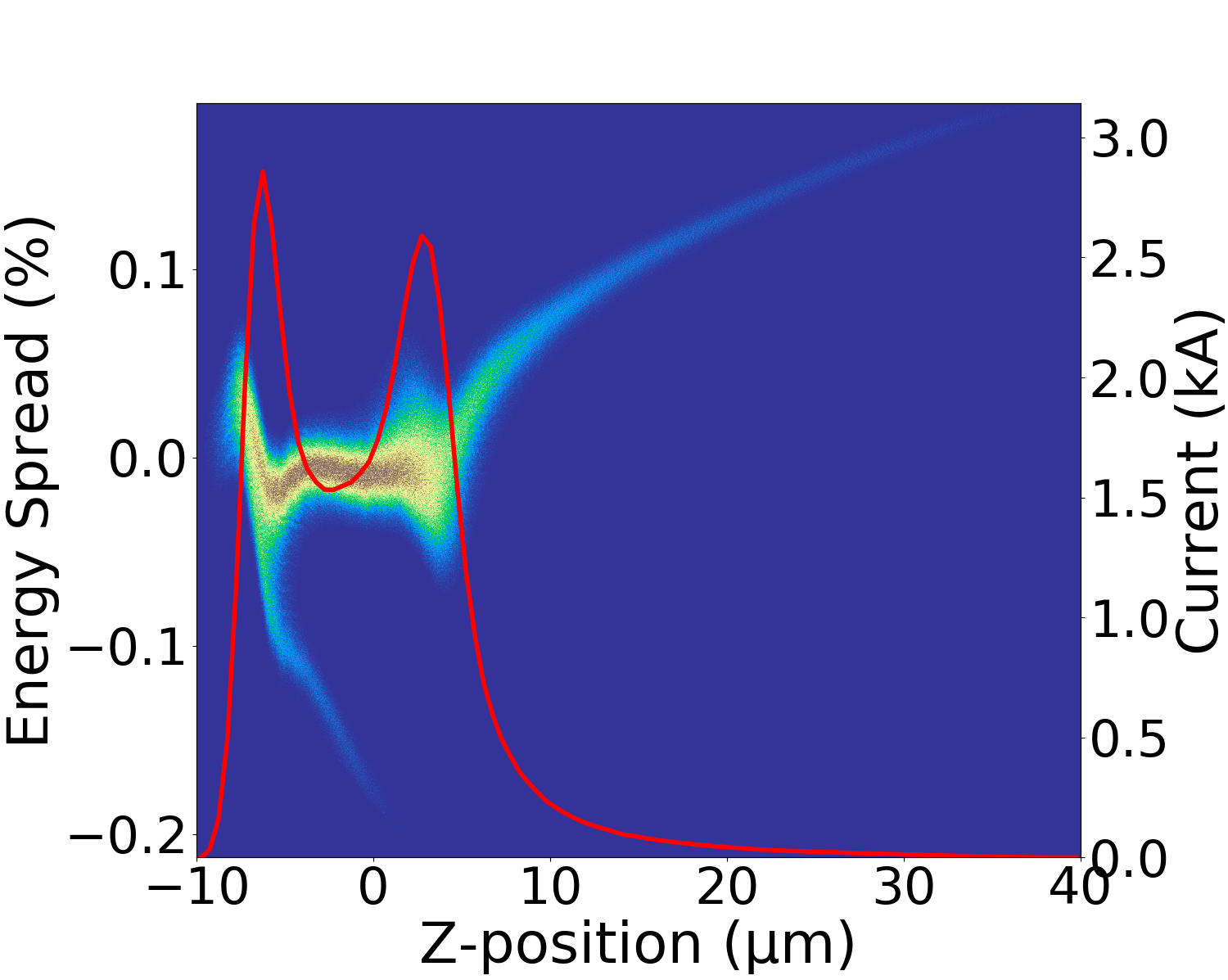}}
	\subfigure[]{
	    \label{slice1}
	    \includegraphics[width=0.3\linewidth]{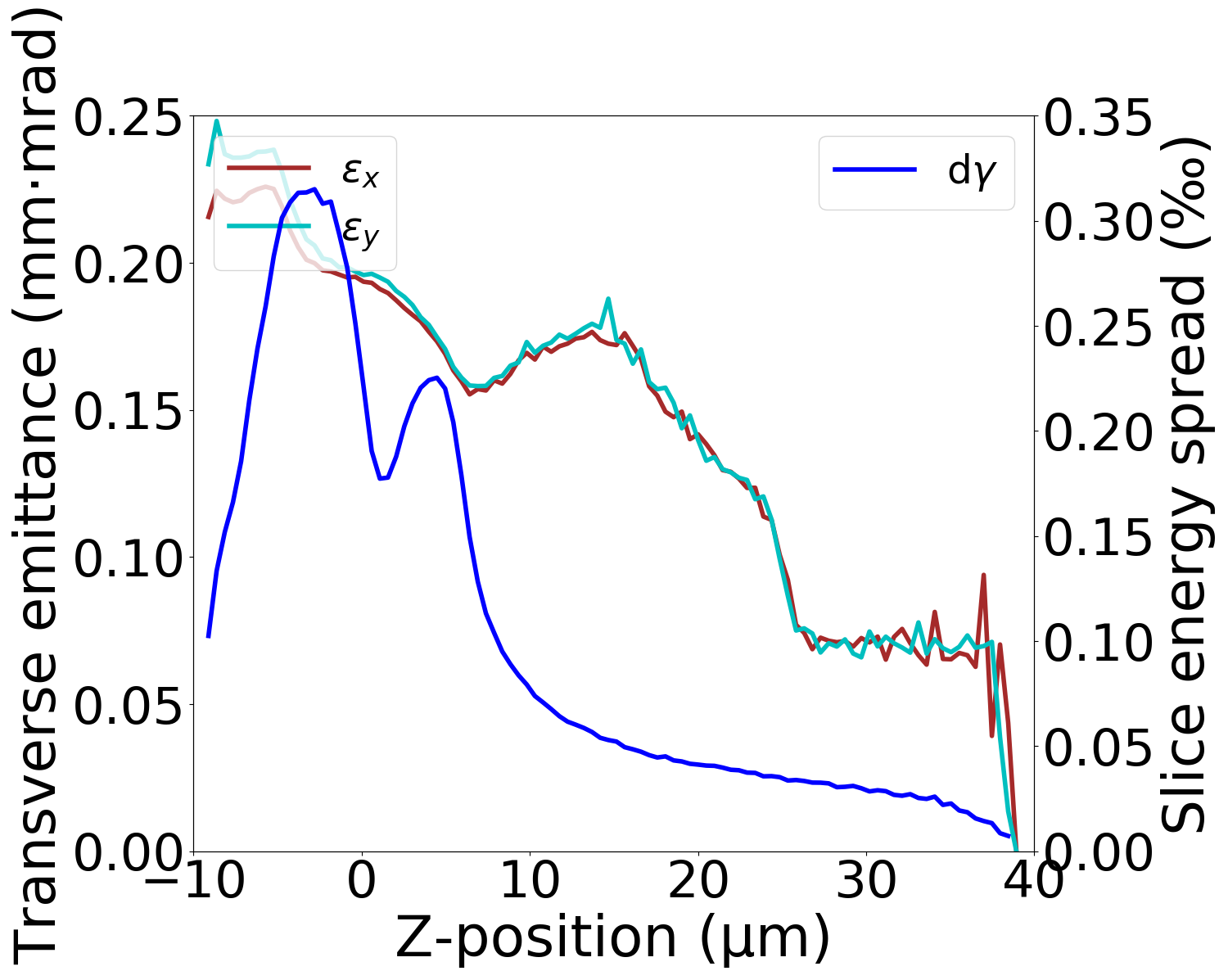}}
	\subfigure[]{
	    \label{mismatch1}
	    \includegraphics[width=0.3\linewidth]{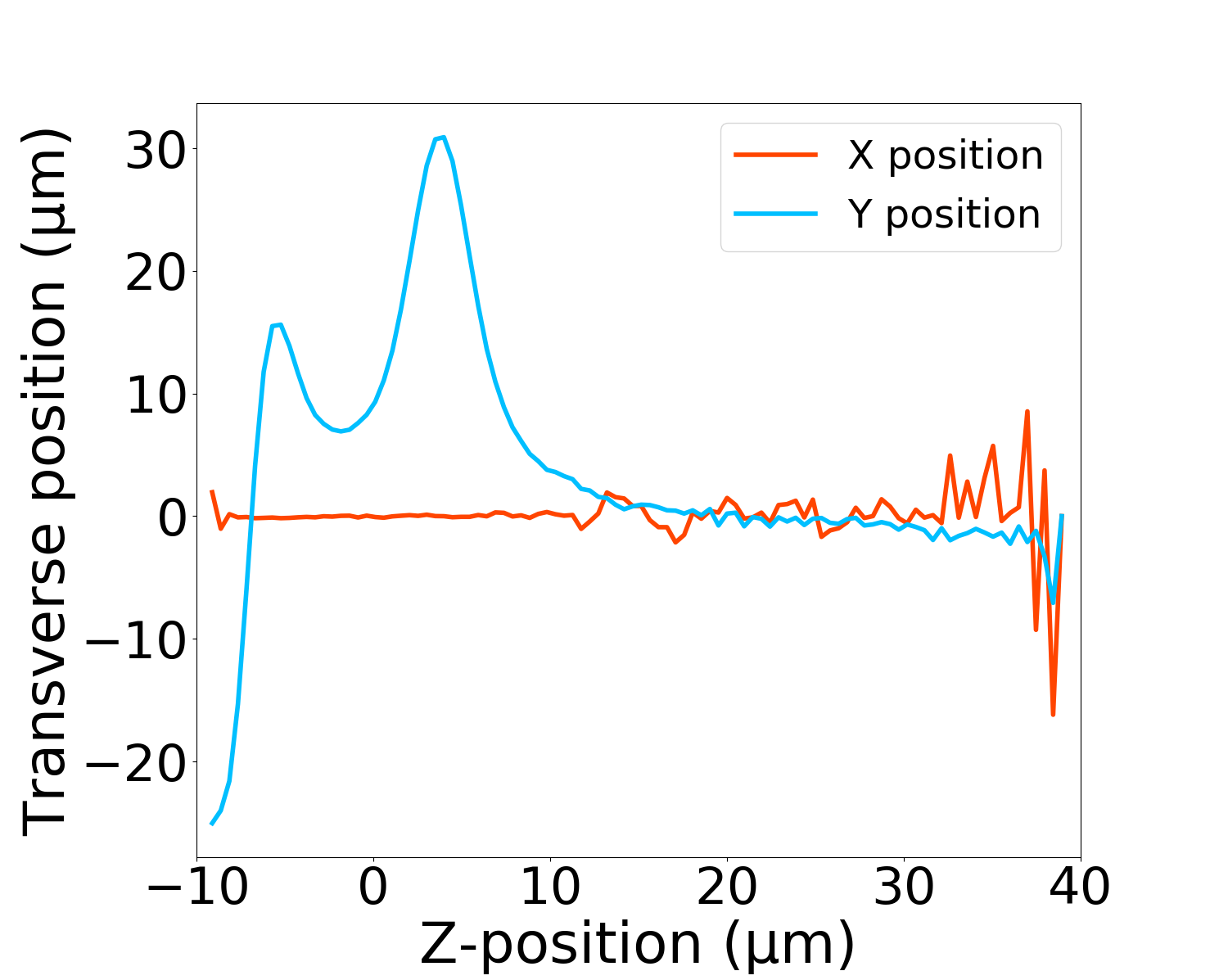}}
	\subfigure[]{
	    \label{lps2}
	    \includegraphics[width=0.3\linewidth]{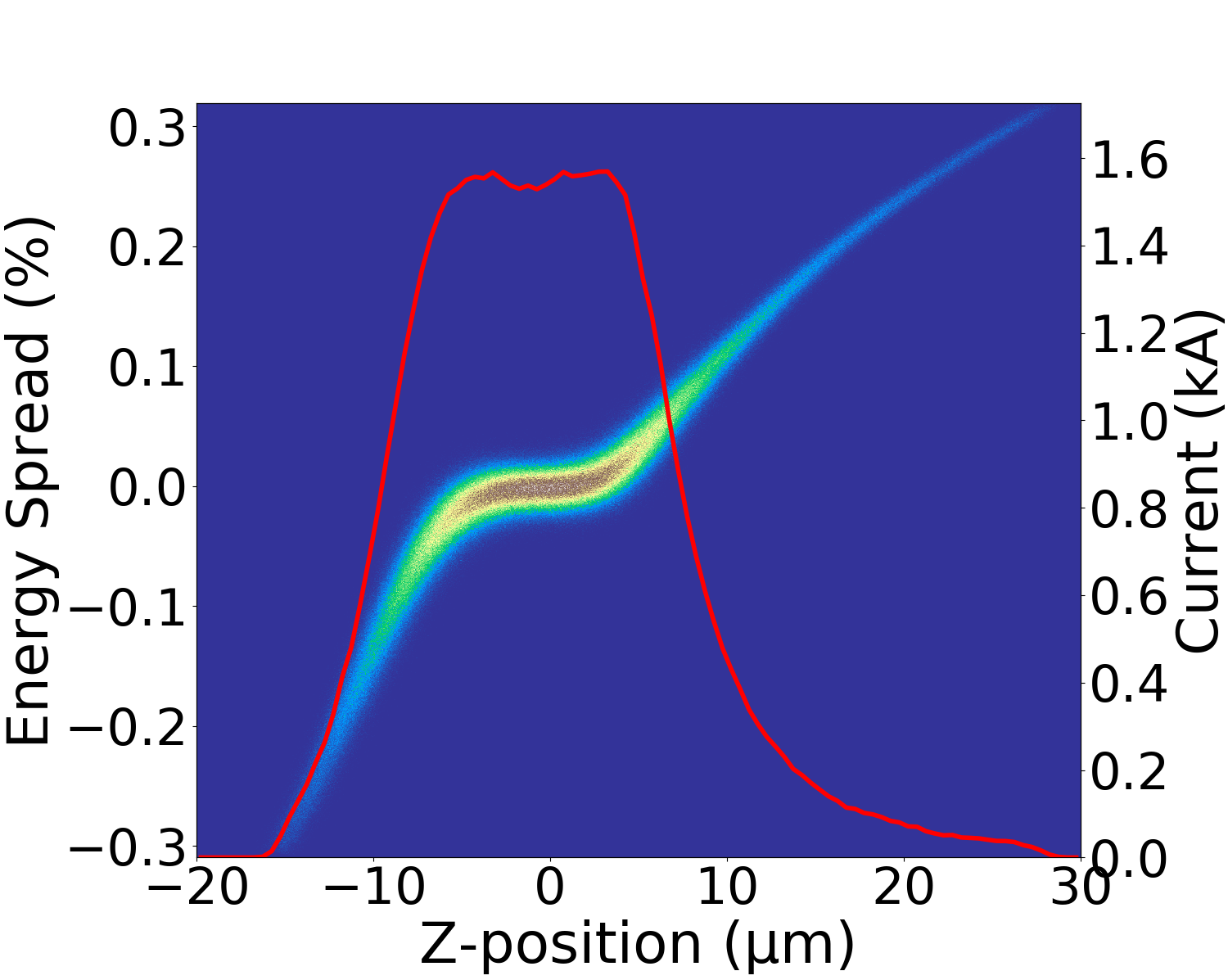}}
	\subfigure[]{
	    \label{slice2}
	    \includegraphics[width=0.3\linewidth]{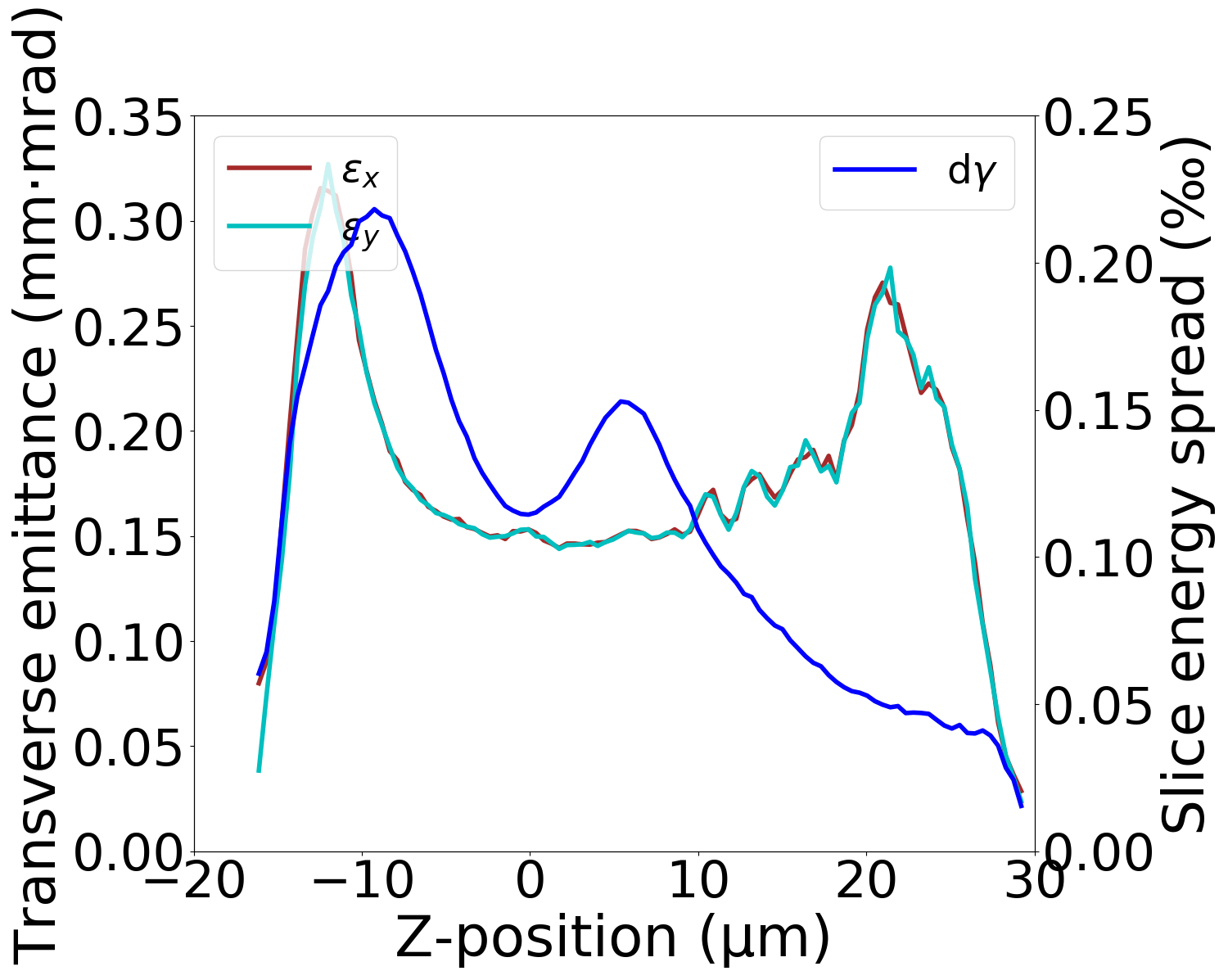}}	 
	\subfigure[]{
	    \label{mismatch2}
	    \includegraphics[width=0.3\linewidth]{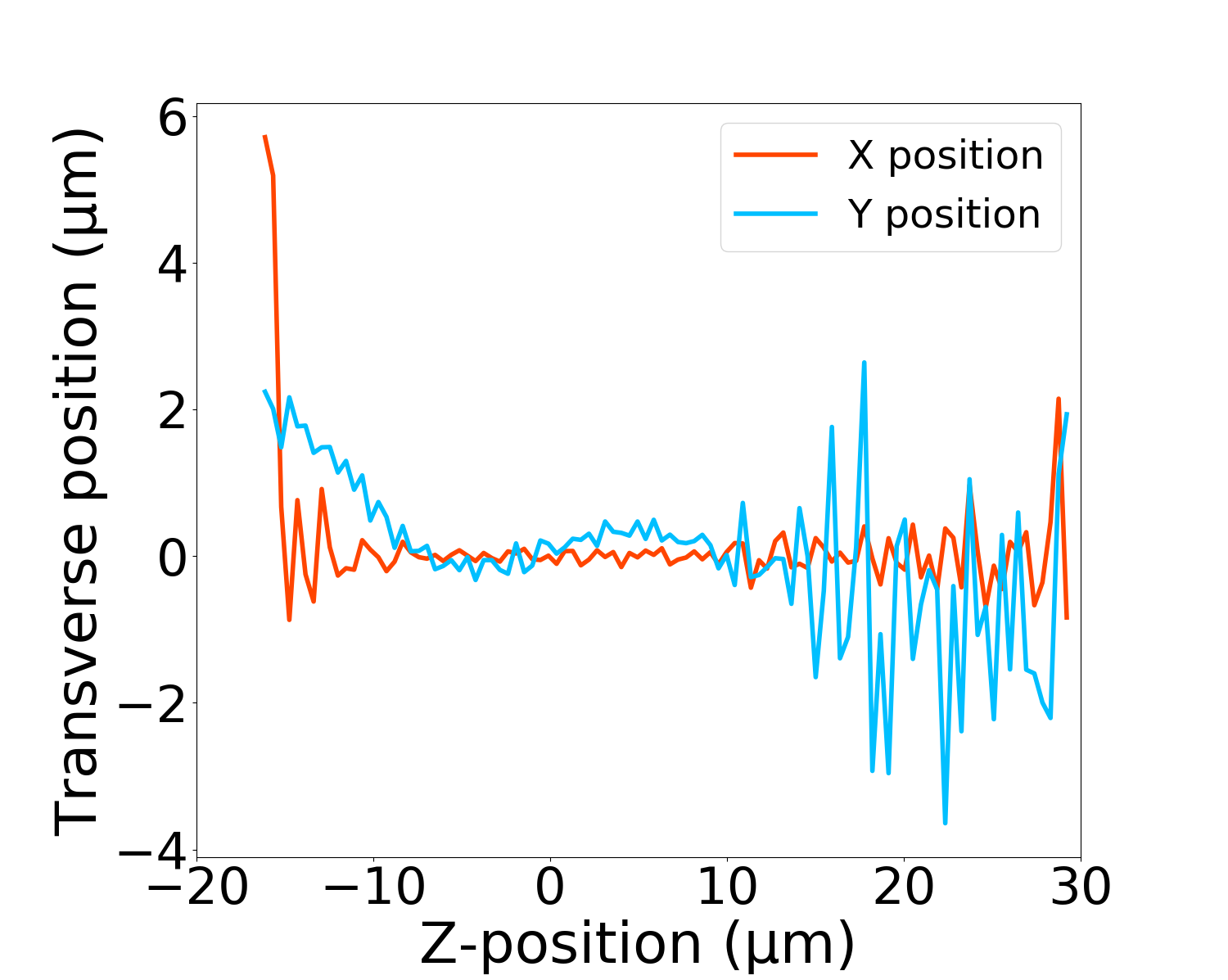}}
	\caption{The comparison of the beam dynamics properties between the previous design (top row) and optimized one with the longitudinal phase space compensation method (bottom row), i.e. detailed longitudinal phase space distribution with the current profile (left column), the corresponding slice transverse emittance and energy spread along the bunch (middle column), and slice beam position in the two dimensions transversely (right column).}
	\label{result_comparison}
\end{figure*}

Fig. \ref{result_comparison} shows the simulation results tracked by ELEGANT software toolkit compared with the previous design \cite{yan2019multi}. The beam energy is 8 GeV and the detailed longitudinal phase space distribution with the current profile of the bunch is plotted in the left column, the slice beam properties of transverse emittance and energy spread are presented in the middle column, and the slice transverse position along the bunch is presented in the right column. The beam longitudinal phase space improvement can be observed in Fig. \ref{lps2}, demonstrating the mitigation of the current horn with a more flat shape in the current profile and energy modulation. The peak current of the optimized beam is nearly 1.6 kA with a flat-top beam core of about 35 femtoseconds. Furthermore, the slice emittance, which is usually regarded as the crucial transverse property in the FEL lasing process, is also kept around 0.15 mm·mrad in the core of the bunch, with the slice energy spread of around $0.015\%$, as shown in Fig. \ref{slice2}. 

\begin{table}[h!]
	\begin{center}
		\caption{Parameters of the two-stage bunch compressors at SHINE.}
		\label{table2}
		\begin{tabular}{l l l l}
    		\toprule
			\textbf{Parameter} & \textbf{BC1} & \textbf{BC2}& \\
			\midrule
			Nominal beam energy (MeV)& 292& 2074\\
			Energy spread ($\%$)&  1.84 & 0.78\\
			$R_{56}$ (mm) &-58.30  &-40.51\\
			Drift length between $1^{st}$ \& $2^{nd}$ ($3^{rd}$ \& $4^{th}$) (m)& 4.71 & 9.91\\
			Drift length between $2^{nd}$ \& $3^{rd}$ (m)& 1.75 & 1.75\\
			Length of chicane dipoles (m)& 0.20 & 0.55\\
			Bending angle (degree) & 4.443 & 2.530\\
	         \bottomrule
		\end{tabular}
	\end{center}
\end{table}	

Moreover, Fig. \ref{mismatch1} demonstrates that the current horn will cause unwanted transverse kicks which are not uniform along the bunch longitudinally. This conspicuous centroid mismatch will smear the beam transverse phase space and lead to the projected emittance growth that ruins the control of the FEL pulse energy and bandwidth. With the longitudinal phase space compensation method utilized, there is no additional transverse yaw under the flat-top distribution, as shown in Fig. \ref{mismatch2}. As a result, the optimized electron beam can significantly improve the FEL performance in the undulator section, which will be presented in the following subsection.

\subsection{FEL performance of EEHG-HGHG cascading operation}
\label{s32}
Normal FEL operations, such as SASE, self-seeding, and externally seeded FELs, can benefit from such a flat-top electron beam. Here, the parameters of the second undulator line of SHINE (FEL-II) are taken as an example to demonstrate the FEL performance improvement. The baseline operation modes of the FEL-II are externally seeded FELs. The first stage of the FEL-II comprises two seed lasers with a wavelength of 270 nm and a pulse length of 20 fs
(FWHM), two modulators of period 240 mm, two magnetic chicanes for the beam manipulation of the echo-enabled harmonic generation (EEHG) \cite{stupakov2009using, xiang2009echo}, and one radiator of period 68 mm. The lengths of the first and second modulators are 3 m and 1.5 m, respectively. The radiator length of the first stage is 20 m. The second stage of the FEL-II comprises one fresh bunch chicane, one modulator of period 68 mm, one dispersion chicane, and one radiator of period 68 mm for cascading operation based on the high gain harmonic generation (HGHG) \cite{yu1991generation}. The radiator length of the second stage is 30 m. 

\begin{figure}[!htb]
	\centering

	\subfigure[]{\includegraphics[width=0.45\linewidth]{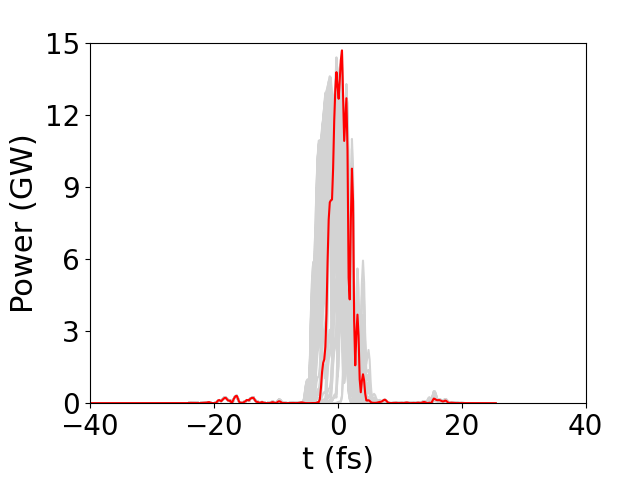}}	
	\subfigure[]{\includegraphics[width=0.45\linewidth]{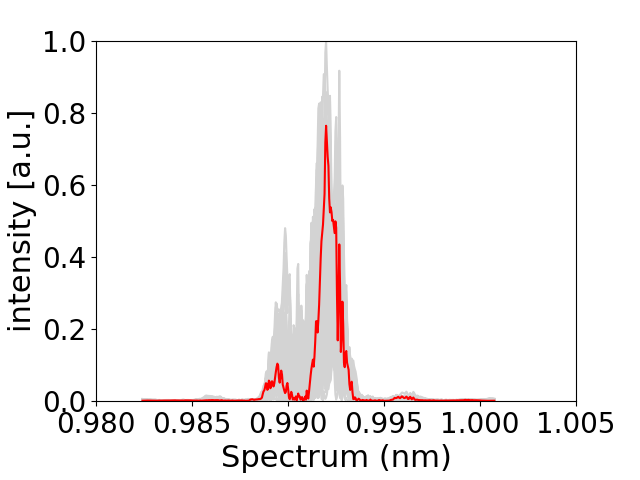}} 
	\subfigure[]{\includegraphics[width=0.45\linewidth]{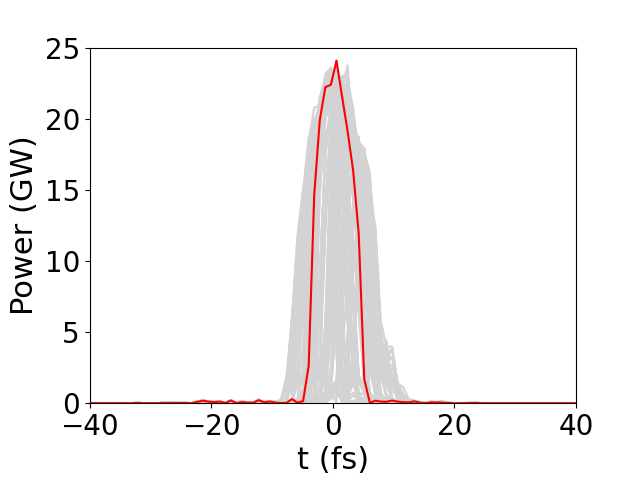}}	  
	\subfigure[]{\includegraphics[width=0.45\linewidth]{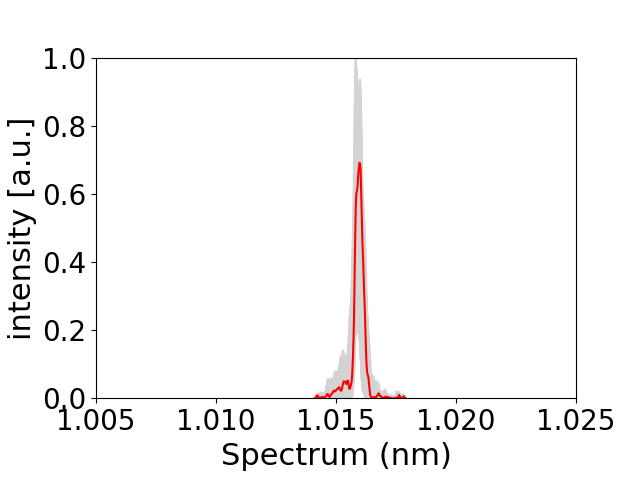}}	
	\caption{Simulated FEL power profile (left) and spectrum (right) of EEHG-HGHG cascading operation based on the previously designed (top) and the optimized electron beams (bottom). The grey lines refer to single shots with different timing jitter between the electron bunch and seed lasers. The red line indicates the case with maximum peak power. }
	\label{FEL}
\end{figure}

The start-to-end simulations of a $\rm 50\times5$ EEHG-HGHG cascading setup are performed with GENESIS1.3 \cite{reiche1999genesis} based on the previously designed electron beam and the optimized electron beam which is illustrated in the previous subsection. In the simulation, the resonance of the radiator of the first stage is set to the 50th harmonic of the seed lasers. The resonance of the radiator of the second stage is set to the 5th harmonic of the first stage. Since the timing jitter between the electron beam and the seed laser is crucial for the EEHG-HGHG cascading operation, 100 relative time jitters with an RMS of 3 fs are considered in the simulations. Fig. \ref{FEL} presents the simulated FEL power profile and spectrum based on the two electron beams. These results show that the FEL performances of the optimized electron beam are significantly better than that of the previous design. Benefiting from a uniform current distribution and a smaller slice energy spread, the optimized beam can be used to generate FEL pulses with higher peak power and a narrower spectrum. The averaged pulse energies obtained based on the previously designed and optimized electron beam are 47 and 148 $\rm \mu J$, respectively. 

\section{Discussion and Conclusion}  
\label{s4}
The final beam dynamics simulation results demonstrate the efficiency of the longitudinal phase space compensation method in the photoinjector. This approach is implemented by feeding the third harmonic RF power to the 1.3 GHz normal buncher downstream of the VHF gun cavity. This scheme can not only compensate the nonlinear energy modulation to adjust the high-order energy spread distribution but also manipulate the beam longitudinal distribution to be more symmetric. This manipulation capability can be demonstrated in Fig. \ref{skew}. The bunch longitudinal distribution can be manipulated from the positive skewness (Fig. \ref{skew1}) to the negative skewness (Fig. \ref{skew3}) through adjusting the RF parameters settings of the two independent modes in the dual-mode buncher. This versatile technique can be be applied in generation of electron bunch with the linearly ramped current profile in the plasma or dielectric wakefield accelerators \cite{england2008generation,roussel2020longitudinal,piot2012generation}.

The nonlinearity in the longitudinal modulation reflects in the different local compressing ratios along the bunch. These electrons experiencing overcompression will overlap with other ones and dramatically enhance the local charge density. Based on it, the current spike formation is analogous to the caustic phenomenon in the optics, and the fundamental mechanism has been studied analytically, readers with interest are kindly referred to \cite{charles2017current,charles2018applications,charles2016caustic}.

As the complicated beam dynamics mentioned above, this photoinjector physical design involves more variables and objectives which need elaborate optimization. The many-objective optimization algorithm NSGA-III is applied in the injector beam dynamics design for the first time to optimize the four beam properties simultaneously, as is presented in Section \ref{s22}. It is verified that it can improve efficiency and provide valuable guidance to further research on the relationship between the optimization objectives.

In conclusion, the nonlinear compression is a common issue for peak current achievement for driving the FEL lasing and is usually vulnerable to current horn formation, which induces the inevitable energy modulation and transverse mismatch that degrade the FEL energy and bandwidth. In order to improve the FEL performance, we focus on the improvement of longitudinal phase space in the photoinjector section where the beam is dominated by the strong space charge repulsion. With the dual-mode buncher deployed, the more flat-top current profile can be achieved with the peak current of 1.6 $\rm kA$ in the bunch charge of 100 $\rm pC$. Furthermore, the start-to-end simulation demonstrated the proposed method could significantly improve FEL performance. It provides an effective method for the correction of nonlinear longitudinal phase space distribution in a high-repetition-rate XFEL facility. Additionally, this technique can be applied to shape the beam longitudinal distribution profile for driving the plasma-wakefield accelerators.

\section{Acknowledgments}
This work was supported by the Youth Innovation Promotion Association CAS (2021282) and the National Natural Science Foundation of China (Nos. 11775294 and 11905275).

\bibliographystyle{elsarticle-num}
\bibliography{ref}
\end{document}